%% file: main.tex
\documentclass[conference]{IEEEtran}
\ifCLASSINFOpdf
\else
\fi
\input{settings}
\usepackage[english]{babel}
\newcounter{bug}
\renewcommand*{\thebug}{\textbf{B\arabic{bug}}}
\newenvironment{bug_ns}[1][]{\refstepcounter{bug}\par\smallskip
\noindent\textbf{Bug \thebug#1}\rmfamily}

\newcounter{vul}
\renewcommand*{\thevul}{\textbf{V\arabic{vul}}}
\newenvironment{vul_ns}[1][]{\refstepcounter{vul}\par\smallskip
\noindent\textbf{Vulnerability \thevul#1}\rmfamily}

\begin{document}
%
\title{\ourtool{}: Reusing Tests for Processor Fuzzing with Contextual Bandits}


\author{
{\rm Chen Chen$^{\dagger}$, Zaiyan Xu$^{\dagger}$, Mohamadreza Rostami$^\ddagger$, David Liu$^{\dagger}$, }\\
{\rm Dileep Kalathil$^{\dagger}$, Ahmad-Reza Sadeghi$^\ddagger$, and Jeyavijayan (JV) Rajendran$^\dagger$}\\
$^\dagger$Texas A\&M University, USA,
$^\ddagger$Technische Universit\"at Darmstadt, Germany\\
} 

\maketitle
\pagestyle{plain}

\input{draft/abstract}


\IEEEpeerreviewmaketitle

\input{draft/introduction}
\input{draft/background}
\input{draft/motivation}
\input{draft/method}

\input{draft/evaluation}
\input{draft/relatedwork}
\input{draft/conclusion}

\input{draft/ethic}

\bibliographystyle{IEEEtran}
\bibliography{main}

\input{draft/appendix}
\end{document}

%% file: settings.tex
\usepackage[nobiblatex]{xurl} 
\usepackage{cite}
\usepackage{amsmath,amssymb,amsfonts, amsthm}
\usepackage{algorithm}
\usepackage{titlesec}
\usepackage[noend]{algpseudocode}
\usepackage{multirow}
\usepackage{graphicx}
\usepackage{textcomp}
\usepackage{xcolor}
\usepackage{mathtools}
\usepackage{xhfill}
\usepackage{lipsum}  
\usepackage{enumitem}
\usepackage{listings}
\usepackage{tikz}
\usepackage[normalem]{ulem}  
\input{listing_style}

\usepackage[most]{tcolorbox}

\definecolor{check_mark_green}{RGB}{43, 103, 22}
\definecolor{default_green}{RGB}{116, 218, 80}
\definecolor{default_gray}{RGB}{128,128,128}

\def\BibTeX{{\rm B\kern-.05em{\sc i\kern-.025em b}\kern-.08em
    T\kern-.1667em\lower.7ex\hbox{E}\kern-.125emX}}

\usepackage{tikz}
\newcommand{\tikzxmark}{%
\tikz[scale=0.23] {
    \draw[red,line width=0.7,line cap=round] (0,0) to [bend left=6] (1,1);
    \draw[red,line width=0.7,line cap=round] (0.2,0.95) to [bend right=3] (0.8,0.05);
}}
\newcommand{\tikzcmark}{%
\tikz[scale=0.23] {
    \draw[check_mark_green,line width=0.7,line cap=round] (0.25,0) to [bend left=10] (1,1);
    \draw[check_mark_green,line width=0.8,line cap=round] (0,0.35) to [bend right=1] (0.23,0);
}}

\usetikzlibrary{fit,calc,backgrounds}
\newcommand*{\tikzmk}[1]{\tikz[remember picture,overlay,] \node (#1) {};\ignorespaces}
\newcommand{\boxitonept}[1]{\tikz[remember picture,overlay]{\node[yshift=-10pt,xshift=-160pt,fill=#1,opacity=0.25,fit={(A)($(B)+(1.6\linewidth,1.1\baselineskip)$)}] {};}\ignorespaces}

\colorlet{pink}{red!40}
\newcommand{\circleicon}[1]{%
    \tikz[baseline=(char.base)] \node[draw, circle, fill=white, text=black, inner sep=0.8pt] (char) {\scriptsize #1};%
}

\newcommand{\ourtool}{\texttt{ReFuzz}}

\newcommand{\mabfuzz}{\texttt{MABFuzz}}
\newcommand{\thehuzz}{\texttt{TheHuzz}}

\newcommand{\morfuzz}{\texttt{MorFuzz}}
\newcommand{\cascade}{\texttt{Cascade}}
\newcommand{\rfuzz}{\texttt{RFUZZ}}
\newcommand{\difuzz}{\texttt{DIFUZZRTL}}
\newcommand{\hypfuzz}{\texttt{HyPFuzz}}

\newcommand{\mints}{\texttt{MINTS}}
\newcommand{\profuzz}{\texttt{ProcessorFuzz}}
\newcommand{\genhuzz}{\texttt{GenHuzz}}

\newcommand{\rc}{{\tt Rocket Core}}
\newcommand{\boomt}{{\tt BOOMV3}}
\newcommand{\boomf}{{\tt BOOMV4}}
\newcommand{\cva}{{\tt CVA6}}
\newcommand{\rsd}{{\tt RSD}}
\newcommand{\picorv}{{\tt PicoRV32}}
\newcommand{\kronos}{{\tt Kronos}}
\newcommand{\vcs}{{\tt VCS}}
\newcommand{\xce}{{\tt Xcelium}}

\newcommand{\riscv}{{\tt RISC-V}}

\newcommand{\arm}{{\tt ARM}}

\theoremstyle{definition}
\newtheorem{defn}{\textbf{Definition}}
\theoremstyle{definition}

\newcommand{\thehuzzrc}{$99.17\%$}
\newcommand{\avereduction}{$98.76\%$}
\newcommand{\aveinteretreduction}{$65.89\%$}
\newcommand{\mintot}{$1557$}

\newcommand{\totmintesthr}{$1.14$}
\newcommand{\trainhorizon}{$10000$}
\newcommand{\trainarm}{$100$}
\newcommand{\trainwindow}{$3$}

\newcommand{\finetunetimehr}{1.30}

\newcommand{\testminitime}{944.28}

\newcommand{\cbhuzzcvainc}{$6.11$}

\newcommand{\cbhuzzrcinc}{$9.33$}

\newcommand{\cbhuzzboomfinc}{$5.94$}
\newcommand{\cbhuzzboomfspdp}{$1818.27$}

\newcommand{\thehuzzrsd}{$68.54$}
\newcommand{\cbhuzzrsd}{$68.45$}

\newcommand{\cbcascadecvainc}{$5.48$}
\newcommand{\cbcascadecvaspdp}{$1086.98$}

\newcommand{\cbcascadercinc}{$1.30$}

\newcommand{\cbcascadeboomfinc}{$0.63$}
\newcommand{\cbcascadeboomfspdp}{$10.90$}

\newcommand{\cascadersd}{$66.91$}
\newcommand{\cbcascadersd}{$67.88$}

\newcommand{\cbcascadersdspdp}{$215.06$}

\newcommand{\cbavetestspdp}{$511.23$}
\newcommand{\cbavetestincov}{$1.89$}

\newcommand{\cbaveallcov}{$3.48$}

\newcommand{\cbavetestcondspdp}{$223.57$}
\newcommand{\cbavetestcondincov}{$2.2$}
\newcommand{\cbaveallcondspdp}{$279.44$}
\newcommand{\cbaveallcondcov}{$3.05$}

\newcommand{\thehuzzvoavetest}{83.33}
\newcommand{\thehuzzvoavetime}{137.44}
\newcommand{\cbthehuzzvoavetest}{4.00}
\newcommand{\cbthehuzzvoavetime}{15.73}
\newcommand{\cbthehuzzvospdptest}{20.83}
\newcommand{\cbthehuzzvospdptime}{8.74}

\newcommand{\cascadevoavetest}{4.67}
\newcommand{\cascadevoavetime}{202.78}
\newcommand{\cbcascadevoavetest}{4.00}
\newcommand{\cbcascadevoavetime}{13.74}
\newcommand{\cbcascadevospduptest}{1.17}
\newcommand{\cbcascadevospdptime}{14.76}

\newcommand{\thehuzzvtwavetest}{302.33}
\newcommand{\thehuzzvtwavetime}{478.43}
\newcommand{\cbthehuzzvtwavetest}{478.33}
\newcommand{\cbthehuzzvtwavetime}{853.65}
\newcommand{\cbthehuzzvtwspdptest}{0.63}
\newcommand{\cbthehuzzvtwspdptime}{0.56}

\newcommand{\thehuzzvtavetest}{375.00}
\newcommand{\thehuzzvtavetime}{596.06}
\newcommand{\cbthehuzzvtavetest}{296.00}
\newcommand{\cbthehuzzvtavetime}{545.18}
\newcommand{\cbthehuzzvtspdptest}{1.27}
\newcommand{\cbthehuzzvtspdptime}{1.09}

\newcommand{\vulalltestspdp}{5.98}
\newcommand{\vulalltimespdp}{6.29}

\newcommand{\tcov}{total coverage}
\newcommand{\spd}{coverage speed}
\newcommand{\scov}{standalone coverage}
\newcommand{\icov}{incremental coverage}
\newcommand{\covi}{coverage increment}

\newcommand{\ccontext}{coverage context}

\newcommand{\vlist}{vulnerability list}
\newcommand{\clist}{coverage list}
\def\removesen{0}
\def\adden{0}
\def\MRen{1}
\def\termdef{1}

\if 1\removesen
\newcommand\red[1]{\textcolor{red!40!white}{\sout{#1}}}
\else
\newcommand\red[1]{}
\fi

\if 1\adden
\newcommand\blue[1]{\textcolor{blue}{#1}}
\else
\newcommand\blue[1]{{#1}}
\fi

\if 1\MRen 
\newcommand\mr[1]{\textcolor{blue}{\textbf{MR:} #1}}
\else
\newcommand\mr[1]{}
\fi

\if 1\termdef
\newcommand\td[1]{\textcolor{orange}{\textbf{term:} #1}}
\else
\newcommand\td[1]{}
\fi


%% file: listing_style.tex
\definecolor{codegreen}{rgb}{0,0.6,0}
\definecolor{codegray}{rgb}{0.5,0.5,0.5}
\definecolor{codepurple}{rgb}{0.58,0,0.82}
\definecolor{backcolour}{rgb}{0.95,0.95,0.92}

\let\othelstnumber=\thelstnumber
\def\createlinenumber#1#2{
    \edef\thelstnumber{%
        \unexpanded{%
            \ifnum#1=\value{lstnumber}\relax
              #2%
            \else}%
        \expandafter\unexpanded\expandafter{\thelstnumber\othelstnumber\fi}%
    }
    \ifx\othelstnumber=\relax\else
      \let\othelstnumber\relax
    \fi
}

\lstdefinestyle{customc}{
  belowcaptionskip=1\baselineskip,
  breaklines=true,
  frame=single,
  xleftmargin=0.35cm,
  xrightmargin=0.15cm,
  numbers=left,
  numbersep=5pt,  
  language=C,
  showstringspaces=false,
  basicstyle=\footnotesize\ttfamily,
  keywordstyle=\bfseries\color{green!40!black},
  commentstyle=\itshape\color{purple!40!black},
  identifierstyle=\color{blue},
  stringstyle=\color{orange},
}

\lstdefinestyle{customcArianeExploit1}{
  breaklines=true,
  frame=single,
  xleftmargin=0.4cm,
  xrightmargin=0.2cm,
  numbers=left,
  numbersep=5pt,  
  language=C,
  showstringspaces=false,
  basicstyle=\footnotesize\ttfamily,
  keywordstyle=\bfseries\color{green!40!black},
  commentstyle=\itshape\color{purple!60!black},
  identifierstyle=\color{blue},
  stringstyle=\color{yellow!50!black},
  morekeywords={asm},
  keywordstyle=[2]\bfseries\color{brown!60!black},
}

\lstdefinestyle{customcArianeExploit}{
  breaklines=true,
  frame=single,
  xleftmargin=0.4cm,
  xrightmargin=0.2cm,
  numbers=left,
  numbersep=5pt,  
  language=C,
  showstringspaces=false,
  basicstyle=\footnotesize\ttfamily,
  keywordstyle=\bfseries\color{blue},
  commentstyle=\itshape\color{green!50!black},
  identifierstyle=\color{black},
  stringstyle=\color{brown},
  morekeywords={asm},
  keywordstyle=[2]\bfseries\color{black},
}

\lstdefinestyle{customlog}{
  breaklines=true,
  frame=single,
  xleftmargin=0.35cm,
  xrightmargin=0.15cm,
  numbers=left,
  numbersep=5pt,  
  language=C,
  showstringspaces=false,
  basicstyle=\footnotesize\ttfamily,
  keywordstyle=\color{blue},
  commentstyle=\itshape\color{purple!40!black},
  identifierstyle=\color{blue},
  stringstyle=\color{orange},
  keywords=[2]{INFO},
  keywords=[3]{ERROR},x
  keywordstyle=[2]\bfseries\color{green!40!black},
  keywordstyle=[3]\bfseries\color{red!500!black},
}

\definecolor{verilogcommentcolor}{RGB}{104,180,104}
\definecolor{verilogkeywordcolor}{RGB}{49,49,255}
\definecolor{verilogsystemcolor}{RGB}{128,0,255}
\definecolor{verilognumbercolor}{RGB}{255,143,102}
\definecolor{verilogstringcolor}{RGB}{160,160,160}
\definecolor{verilogdefinecolor}{RGB}{128,64,0}
\definecolor{verilogoperatorcolor}{RGB}{0,0,128}
\definecolor{pointcolor}{RGB}{192,0,0} 
\lstdefinestyle{prettyverilog}{
   language           = Verilog,
   commentstyle       = \color{verilogcommentcolor},
   alsoletter         = \$'0123456789\`,
   literate           = *{+}{{\verilogColorOperator{+}}}{1}%
                         {-}{{\verilogColorOperator{-}}}{1}%
                         {@}{{\verilogColorOperator{@}}}{1}%
                         {;}{{\verilogColorOperator{;}}}{1}%
                         {*}{{\verilogColorOperator{*}}}{1}%
                         {?}{{\verilogColorOperator{? }}}{1}%
                         {:}{{\verilogColorOperator{:}}}{1}%
                         {<}{{\verilogColorOperator{<}}}{1}%
                         {>}{{\verilogColorOperator{>}}}{1}%
                         {!}{{\verilogColorOperator{!}}}{1}%
                         {^}{{\verilogColorOperator{^}}}{1}%
                         {|}{{\verilogColorOperator{| }}}{1}%
                         {||}{{\verilogColorOperator{|| }}}{1}%
                         {=}{{\verilogColorOperator{= }}}{1}%
                         {==}{{\verilogColorOperator{== }}}{1}%
                         {=>}{{\verilogColorOperator{=> }}}{1}%
                         {[}{{\verilogColorOperator{[}}}{1}%
                         {]}{{\verilogColorOperator{]}}}{1}%
                         {(}{{\verilogColorOperator{(}}}{1}%
                         {)}{{\verilogColorOperator{)}}}{1}%
                         {,}{{\verilogColorOperator{,}}}{1}%
                         {.}{{\verilogColorOperator{.}}}{1}%
                         {~}{{\verilogColorOperator{$\sim$}}}{1}%
                         {\%}{{\verilogColorOperator{\%}}}{1}%
                         {\&}{{\verilogColorOperator{\& }}}{1}%
                         {\&\&}{{\verilogColorOperator{\&\& }}}{1}%
                         {\#}{{\verilogColorOperator{\#}}}{1}%
                         {\ /\ }{{\verilogColorOperator{\ /\ }}}{3}%
                         {\ _}{\ \_}{2}%
                        ,
   morestring         = [s][\color{verilogstringcolor}]{"}{"},%
   identifierstyle    = \color{black},
   vlogdefinestyle    = \color{verilogdefinecolor},
   vlogconstantstyle  = \color{verilognumbercolor},
   vlogsystemstyle    = \color{verilogsystemcolor},
   basicstyle         = \scriptsize\fontencoding{T1}\ttfamily,
  columns=fullflexible, 
   keywordstyle       = \bfseries\color{verilogkeywordcolor},
   morekeywords      = {val, when, port, coverage, unique},
   numbers            = left,
   numbersep          = 5pt,
   tabsize            = 2,
   escapeinside       = {/*!}{!*/},
   upquote            = true,
   sensitive          = true,
   showstringspaces   = false, 
   frame              = single,
   breaklines         = true,
   abovecaptionskip   = 0pt,
   belowcaptionskip   = 2pt,   
   xleftmargin        =0.35cm,
   xrightmargin       =0.15cm,
   captionpos         = t,
   emph               = {Point, Point0, Point1, Point2, Point3, Point4, Point5, Point6, Point7, Point8, Point9},
   emphstyle          =\color{pointcolor},
   emph               = {[2] STVEC,SCOUNTEREN,MSTATUS,MTVEC,ML1_ICACHE_MISS,ML1_DCACHE_MISS,MITLB_MISS,MDTLB_MISS,
                             MLOAD,MSTORE,MEXCEPTION,MEXCEPTION_RET,MBRANCH_JUMP,MCALL,MRET,MMIS_PREDICT,MSB_FULL,
                             MIF_EMPTY,MHPM_COUNTER_17,MHPM_COUNTER_18,MHPM_COUNTER_19,MHPM_COUNTER_20,MHPM_COUNTER_21,
                             MHPM_COUNTER_22,MHPM_COUNTER_23,MHPM_COUNTER_24,MHPM_COUNTER_25,MHPM_COUNTER_26,MHPM_COUNTER_27,
                             MHPM_COUNTER_28,MHPM_COUNTER_29,MHPM_COUNTER_30,MHPM_COUNTER_31}, 
   emphstyle          = {[2]\bfseries\color{verilogkeywordcolor}}
}

\makeatletter

\newcommand\language@verilog{Verilog}
\expandafter\lst@NormedDef\expandafter\languageNormedDefd@verilog%
  \expandafter{\language@verilog}
  
\lst@SaveOutputDef{`'}\quotesngl@verilog
\lst@SaveOutputDef{``}\backtick@verilog
\lst@SaveOutputDef{`\$}\dollar@verilog

\newcommand\getfirstchar@verilog{}
\newcommand\getfirstchar@@verilog{}
\newcommand\firstchar@verilog{}
\def\getfirstchar@verilog#1{\getfirstchar@@verilog#1\relax}
\def\getfirstchar@@verilog#1#2\relax{\def\firstchar@verilog{#1}}

\newcommand\addedToOutput@verilog{}
\lst@AddToHook{Output}{\addedToOutput@verilog}

\newcommand\constantstyle@verilog{}
\lst@Key{vlogconstantstyle}\relax%
   {\def\constantstyle@verilog{#1}}

\newcommand\definestyle@verilog{}
\lst@Key{vlogdefinestyle}\relax%
   {\def\definestyle@verilog{#1}}

\newcommand\systemstyle@verilog{}
\lst@Key{vlogsystemstyle}\relax%
   {\def\systemstyle@verilog{#1}}

\newcount\currentchar@verilog
  
\newcommand\@ddedToOutput@verilog
{%
   \ifnum\lst@mode=\lst@Pmode%
      \expandafter\getfirstchar@verilog\expandafter{\the\lst@token}%
      \expandafter\ifx\firstchar@verilog\backtick@verilog
         \let\lst@thestyle\definestyle@verilog%
      \else
         \expandafter\ifx\firstchar@verilog\dollar@verilog
            \let\lst@thestyle\systemstyle@verilog%
         \else
            \expandafter\ifx\firstchar@verilog\quotesngl@verilog
               \let\lst@thestyle\constantstyle@verilog%
            \else
               \currentchar@verilog=48
               \loop
                  \expandafter\ifnum%
                  \expandafter`\firstchar@verilog=\currentchar@verilog%
                     \let\lst@thestyle\constantstyle@verilog%
                     \let\iterate\relax%
                  \fi
                  \advance\currentchar@verilog by \@ne%
                  \unless\ifnum\currentchar@verilog>57%
               \repeat%
            \fi
         \fi
      \fi
   \fi
}

\lst@AddToHook{PreInit}{%
  \ifx\lst@language\languageNormedDefd@verilog%
    \let\addedToOutput@verilog\@ddedToOutput@verilog%
  \fi
}

\newcommand{\verilogColorOperator}[1]
{%
  \ifnum\lst@mode=\lst@Pmode\relax%
   {\bfseries\textcolor{verilogoperatorcolor}{#1}}%
  \else
    #1%
  \fi
}

\makeatother

\lstdefinestyle{mystyle}{
    commentstyle=\textit,
    keywordstyle=\textbf,
    stringstyle=\color{codepurple},
    basicstyle=\ttfamily,
    breakatwhitespace=false,         
    breaklines=true,      
    frame=single, 
    framexleftmargin=\parindent,
    captionpos=b,                    
    keepspaces=true,                 
    numbers=left,    
    numberstyle=\normalsize,
    stepnumber=1,
    numbersep=5pt,   
    xleftmargin=1.5\parindent,
    showspaces=false,                
    showstringspaces=false,
    showtabs=false,                  
    tabsize=2
}

%% file: draft/abstract.tex
\begin{abstract}
Processor designs rely on iterative modifications and reuse well-established designs. 
However, this reuse of prior designs also leads to similar vulnerabilities across multiple processors.
As processors grow increasingly complex with iterative modifications, efficiently detecting vulnerabilities from modern processors is critical.
Inspired by software fuzzing, hardware fuzzing has recently demonstrated its effectiveness in detecting processor vulnerabilities. 
Yet, to our best knowledge, existing processor fuzzers fuzz each design individually, lacking the capability to understand known vulnerabilities in prior processors to fine-tune fuzzing to identify similar or new variants of vulnerabilities. 

To address this gap, we present \ourtool{}, an adaptive fuzzing framework that leverages \textit{contextual bandit} to reuse highly effective tests from prior processors to fuzz a processor-under-test (PUT) within a given ISA.
By intelligently \red{tweaking}\blue{mutating} tests that trigger vulnerabilities in prior processors, \ourtool{} effectively detects similar and new variants of vulnerabilities in PUTs. 
\ourtool{} \red{has discovered}\blue{uncovered} three new security vulnerabilities and two new functional bugs.
\blue{\ourtool{} detected one vulnerability by reusing a test that triggers a known vulnerability in a prior processor.}
\red{The first}\blue{One} functional bug exists across three processors that share design modules\red{: \rc{}, \boomt{}, and \boomf{}}.
The second\red{ functional} bug has two variants\red{ on \boomt{} and \boomf{}}. 
Additionally, \ourtool{} reuses highly effective tests to enhance efficiency in coverage\red{ achievement},\red{ thereby} achieving an average $\cbavetestspdp{} \times$ coverage speedup and up to $\cbhuzzrcinc{}\%$ more total coverage, compared to existing fuzzers. 

\end{abstract}

%% file: draft/introduction.tex
\section{Introduction}\label{sec:intro}
Processors, as the core of computing systems, are crucial not only for performance but also for system security.
Over the past 60 years, instruction set architectures (ISAs) have abstracted the functionality of processors independently of their designs.
This abstraction makes the main goal of processor designs\red{ to} enhance performance, dependability, energy efficiency, and fast real-time responses through the introduction of new microarchitectures, rather than adding new functionalities or altering input and output spaces~\cite{patterson2016computer}.

Consequently, processor designs rely on iterative modifications and extensive reuse of well-established designs.
For example, \textit{Intel} extends successful processor generations such as \textit{Tiger Lake} into subsequent generations like \textit{Alder Lake} and \textit{Raptor Lake}~\cite{intel_user_manual,intel_product}.
Moreover, the \textit{design reuse} strategy is typical in hardware.
According to a 2023 worldwide semiconductor survey, over two-thirds of non-memory system-on-chips (SoCs) and integrated circuits (ICs) reuse existing designs~\cite{ipreuse}.
Supporting this trend, hardware programming languages like \textit{Chisel} have been developed to facilitate design reuse~\cite{bachrach2012chisel}.
For example, \boomt{}~\cite{boom} reuses large portions of codebase from another processor, \rc{}~\cite{rocket_chip_generator}. 

While abstraction of functionalities and design reuse \red{provide engineering advantages}\blue{reduce workload}, they also allow latent bugs and vulnerabilities to propagate across processors.
For example, RISC-V processors, \cva{}~\cite{cva6}, \picorv{}~\cite{picorv32}, and \kronos{}~\cite{kronos}, incorrectly raise exceptions for valid \texttt{FENCE} and \texttt{FENCE.I} instructions due to faulty decoding logic~\cite{kande2022thehuzz,solt2024cascade,xu2023morfuzz}.  
Similarly, \cva{} and \boomt{} incorrectly raise exceptions when accessing page table entries (PTE) that violate physical memory attribute (PMA) checks~\cite{xu2023morfuzz}. 

As processor designs continue to grow more complex, efficiently verifying their integrity and security becomes increasingly challenging.
Patching these vulnerabilities post-silicon is costly, as the flaws exist physically within the hardware~\cite{ender2020unpatchable}, often requiring kernel and microcode updates~\cite{kernel_patch,microcode_patch}, disabling microarchitectures~\cite{disable_smt}, or even recalling products~\cite{intel_1994}. 
Such mitigation degrades performance and significantly impacts vendors' finances and reputation.
Detecting vulnerabilities during the pre-silicon stage (i.e., before fabricating the processors) is therefore critical.

\noindent\textbf{Industrial verification flow.} Industry applies both direct and random testing to verify processors before fabrication~\cite{cpu_verif}.
Direct testing uses the internals of processor designs and known vulnerabilities, such as common vulnerabilities and exposures (CVEs) and common weakness enumerations (CWEs), to manually create directed tests~\cite{cwe_test}. 
These tests are usually reused across different designs for vulnerability detection and coverage achievement~\cite{seedreuse}.
Random testing generates instruction sequences to verify processor behaviors. 
However, direct testing requires deep domain knowledge, while random testing struggles to effectively verify large-scale designs~\cite{kande2022thehuzz}.

\noindent\textbf{Processor fuzzing at the pre-silicon stage.} Inspired by the success of software fuzzing, processor fuzzing has emerged as an effective approach for detecting vulnerabilities in modern processors~\cite{rostami2024fuzzerfly,canakci2022processorfuzz,solt2024cascade,hur2021difuzzrtl,kande2022thehuzz,xu2023morfuzz,rajapaksha2023sigfuzz,sugiyama2023surgefuzz}.
Processor fuzzers have proven effective for identifying a wide range of vulnerabilities during the pre-silicon stage, including functional incorrectness~\cite{xu2023morfuzz}, time side-channels~\cite{borkar2024whisperfuzz}, and speculative vulnerabilities~\cite{hur2022specdoctor, specure,de2025phantom,ghaniyoun2021introspectre}.
Typically, processor fuzzers use a set of seeds to initiate fuzzing, with the effectiveness of a seed directly influencing the efficiency of vulnerability detection and coverage achievement~\cite{zhu2022fuzzing}.
Existing research treats processor fuzzing as a more effective alternative to random testing, fuzzing each processor individually, generating seeds from scratch, and improving efficiency through advanced seed generation techniques~\cite{chen2023hypfuzz,chen2023psofuzz,xu2023morfuzz,solt2024cascade,rostami2024chatfuzz,wugenhuzz,wu2025hfl}.

\begin{figure}
    \centering
    \includegraphics[width=0.85\linewidth,trim={30 18 38 24}, clip]{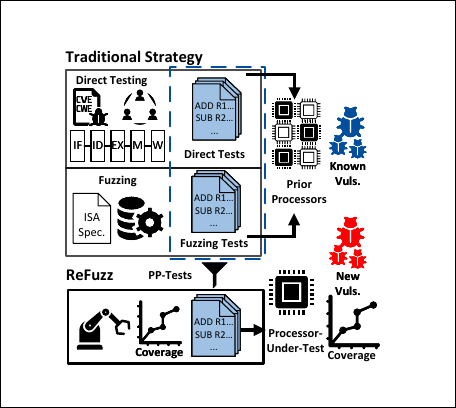}
    \caption{\ourtool{}, a novel fuzzing framework that leverages effective tests from prior processors to enhance fuzzing efficiency on processor-under-tests (PUTs). Vuls. means Vulnerabilities.}
    \label{fig:high-level}
\end{figure}

To our best knowledge, no hardware fuzzer leverages insights from direct testing by reusing tests\footnote{A processor test is a binary executable with a sequence of instructions.} from \textit{prior processors} (PP-tests) to \red{fine-tune fuzzing for vulnerability detection and coverage achievement in processor-under-tests (PUTs)}\blue{guide fuzzing of a processor-under-test (PUT) and enhance coverage and vulnerability detection}\footnote{For \red{the sake of }brevity, we call processors that share the ISA with PUTs as ``prior'' processors and the tests executed on them as ``prior-processor'' tests (PP-tests).}, as shown in Figure~\ref{fig:high-level}.
\blue{Developing such a method is the core contribution of this work and aligns hardware fuzzing with the reuse trend in the hardware development cycle.}
\red{Addressing this research gap and solving its related challenges is our primary contribution in this paper.}

\noindent\textbf{Reusing PP-tests.} The PP-tests can enhance fuzzing efficiency on PUTs for three reasons: 
(i)~Processors within the same ISA mostly share input and output spaces and functionalities, enabling effective reuse of PP-tests.
(ii)~Complex functionalities often remain vulnerable across generations. 
For example, despite the \texttt{FDIV} bug being discovered in \textit{Intel} \textit{Pentium} processors in 1994~\cite{intel_1994}, \textit{AMD} still reports bugs of floating point units in its \textit{Zen} family processors in 2023 due to the complexity of floating-point arithmetic~\cite{amd_revision}. 
Similarly, processor fuzzers like \cascade{}~\cite{solt2024cascade} report multiple vulnerabilities related to the memory synchronization function in RISC-V processors. This function is challenging to implement while maintaining both memory consistency and pipeline efficiency.
(iii)~New microarchitectures can introduce variants of known vulnerabilities that exist in prior processors, making PP-tests valuable starting points for uncovering related flaws.


However, we observe that directly executing PP-tests on PUTs fails to detect variants of known vulnerabilities or to explore the new microarchitectures.
While hardware fuzzing addresses these limitations, it introduces its own challenges:
(i) A fuzzer can over- or under-mutate a PP-test, reducing both the fuzzer’s efficiency and effectiveness. 
The fuzzer must carefully balance between the time to mutate a given test and the need to proceed to the next.
(ii) The test effectiveness varies during fuzzing, requiring dynamic evaluation and prioritization.

\noindent\textbf{\ourtool{}.} To overcome the challenges, we introduce \ourtool{}, the first fuzzing framework that leverages contextual bandit (CB) algorithms to adaptively reuse PP-tests as seeds.
\ourtool{} uses the exploration-exploitation trade-off inherent to CB algorithms to balance between reusing the current test and switching to the next.
\ourtool{} evaluates coverage increment of a PP-test at different total coverage to precisely prioritize seeds, enhancing both coverage and vulnerability detection.

\noindent Overall, the main contributions of this paper are:
\begin{itemize}[align=parleft,leftmargin=*]
    \item We develop the first framework, \ourtool{}, that leverages CB algorithms to guide test reuse from prior processors as seeds for fine-tuning fuzzers on PUTs. 
    Unlike existing approaches that fuzz each PUT independently, \ourtool{} exploits ISA abstraction and common design reuse to provide effective tests across processors following the same ISA.
    \item \ourtool{} is agnostic to any processor fuzzers and random testing that require seeds to initiate processes.
    \item We evaluate \ourtool{} on five widely-used and open-sourced RISC-V processors with diverse microarchitectures and achieve an average $\cbavetestspdp{}\times$ coverage speedup over baseline fuzzers. 
    \ourtool{} also outperforms baseline fuzzers and achieves up to $\cbhuzzrcinc{}\%$ more \tcov{} (see Section~\ref{sec:eva}).
    \item \ourtool{}\red{ successfully} detected three new\red{ security} vulnerabilities and two new functional bugs. 
    \ourtool{} detected one vulnerability by reusing a PP-test that triggers a known vulnerability in a prior processor, resulting in a memory deadlock exploitable for denial-of-service attacks.
    \rc{}~\cite{rocket_chip_generator}, \boomt{}, and \boomf{}~\cite{boom} have the same bug due to reusing the same module.
    \boomt{} and \boomf{} share another\red{ bug}, and \boomf{} has more variants due to its new microarchitectures.
\end{itemize}

%% file: draft/background.tex
\section{Background}\label{sec:bg}


\subsection{Verification of Processor Design}
To manage design complexity, verification of modern processors is conducted at multiple levels. 
At the unit level, verification targets individual components like decoders or adders.
The subsystem level targets integrated groups of modules that perform specific functions like cache coherence. 
At the architecture level, verification ensures that the processor is compliant with its ISA specifications, which is the primary focus of most hardware fuzzers~\cite{cpu_veri_stages}.

To support multi-level verification, both industry and academia employ a comprehensive toolbox of formal and dynamic techniques~\cite{synopsys_toolbox}. 
\textbf{Formal verification}, primarily used at the unit and subsystem levels, relies on predefined assertions to validate functional correctness and security properties. 
At the architecture level, dynamic verification is more common and includes \textbf{random testing}, which generates instruction sequences to explore design behaviors, and \textbf{direct testing}, which applies curated test suites or handcrafted tests based on known vulnerabilities (e.g., CVEs, CWEs)~\cite{riscv_verif}.
Because many processors share similar verification goals, such as covering corner cases and detecting variants of vulnerabilities, directed tests from prior processors are commonly reused across processor generations\red{, significantly reducing} \blue{to reduce} verification effort~\cite{seedreuse}.


\subsection{Hardware Processor Fuzzers}\label{sec:bg_hf}
Fuzzing is a dynamic technique that verifies designs through iterative test generation and execution~\cite{zhu2022fuzzing}.   
Processor fuzzers typically produce instruction sequences based on the instruction set architecture (ISA) of the processor-under-test (PUT). 
A processor fuzzer includes four core components: \textit{seed generator}, \textit{mutation engine}, \textit{feedback engine}, and \textit{vulnerability detector}~\cite{rostami2024fuzzerfly}.
The \textit{seed generator} generates an initial set of tests as \textbf{seeds}, by randomly selecting instruction opcodes and operands~\cite{hur2021difuzzrtl}.
A fuzzer then simulates or emulates these seeds on the PUT and collects feedback data and output used by the \textit{feedback engine} to guide mutations and the \textit{vulnerability detector} to identify bugs. 

The \textit{feedback engine} often employs code coverage as feedback~\cite{kande2022thehuzz}, which monitors the amount of hardware logic explored, such as branch statements, finite-state machines~(FSMs), and toggled bits.
Additionally, hardware fuzzers use customized metrics as feedback, such as control-register coverage~\cite{hur2021difuzzrtl}, which tracks the states of multiplexer signals, and control and status register (CSRs) coverage~\cite{canakci2022processorfuzz}, which monitors the values of CSRs in the PUT.
The \textit{vulnerability detector} identifies vulnerabilities using either \textbf{assertions}, which check if certain conditions hold true at runtime~\cite{fuzzhwlikesw}, or \textbf{differential testing}, which compares the PUT's outputs against a golden-reference model (GRM)~\cite{kande2022thehuzz,hur2021difuzzrtl, xu2023morfuzz,solt2024cascade,wugenhuzz}; mismatches represent potential vulnerabilities in the PUT.

After executing the initial tests, the feedback engine identifies ``interesting tests" that reach new coverage points for further exploration of design spaces.
These tests guide the \textit{mutation engine} to generate new tests.
The mutation engine performs data manipulation, such as bit flips and swaps~\cite{kande2022thehuzz,rfuzz}, similar to the strategies used in the most popular software fuzzer, American Fuzzy Lop (AFL)~\cite{citeafl}. 
Mutation operators may mutate instruction operands or alter entire instructions.
The \textit{seed generator} and \textit{mutation engine} automate test generation, and the effectiveness of seeds is crucial for determining the efficiency and effectiveness of a fuzzing campaign~\cite{zhu2022fuzzing}.


\subsection{Bandit Problems}~\label{sec:bg_bandit}
Bandit problems are a class of reinforcement learning problems focused on action selection without modeling future state transitions.
The objective is to learn a \textbf{policy} that maximizes expected cumulative reward by balancing exploration and exploitation~\cite{lattimore2020bandit}. 

\noindent\textbf{Contextual bandit (CB)} is a type of bandit problem that includes contextual information.
Before taking an action, the learner observes the context of the environment. 
The learner then selects an action and receives a reward only for that action, without feedback on the unchosen alternatives. 
This feature makes CB algorithms particularly suited for real-world scenarios, where environments are dynamic and involve a large number of actions (e.g., thousands)~\cite{lattimore2020bandit}.

In general, CB consists of six core components: (i) agent: the decision maker selecting actions; (ii) environment: the source of context and reward feedback; (iii) set of arms ($\mathcal{A}$): the set of available actions; (iv) context ($c$): information observed before making a decision; (v) policy ($\pi$): the mapping from context to actions; (vi) reward ($r$): feedback from environment for the chosen action.
At each time step $t$, the agent observes a context $c_t$ from the environment, selects an action $a \in A$ according to policy $\pi(\cdot|c_t)$, and receives a reward $r_t = f(c_t, a)$ from the environment.

%% file: draft/motivation.tex
\section{Observations on Reusing and Mutating Tests}\label{sec:motivation}

Directly reusing prior-processor tests (PP-tests) often falls short when detecting variants of vulnerabilities or achieving comprehensive coverage on processor-under-tests (PUTs). 
While PP-tests remain valuable for their original verification purposes, they exhibit inherent limitations in identifying variants of vulnerabilities that manifest uniquely within evolved microarchitectures.
In this section, we analyze the limitations of directly using PP-tests to detect variants of vulnerabilities or enhance coverage on PUTs and show how processor fuzzers can overcome these limitations by mutating tests intelligently.


\subsection{Case Study: Detecting Variants of Vulnerabilities}\label{sec:casestudy}

\ourtool{} detects a new bug affecting both \boomt{} and \boomf{}~\cite{boom}, where improper updates to the CSRs cause certain instructions observed to increment the committed instruction counter (\texttt{minstret}) by two.
This behavior violates the RISC-V specification, which mandates that each committed instruction increments \texttt{minstret} by one~\cite{riscvisa}.
Accurate accounting of committed instructions is essential for performance profiling~\cite{demme_bottlenecks}, bug reproduction~\cite{weaver_debugging}, and anomaly detection~\cite{wang_performance_evaluation}.

\begin{figure}
    \centering
    \includegraphics[width=0.95\linewidth,trim={24 20 25 22}, clip]{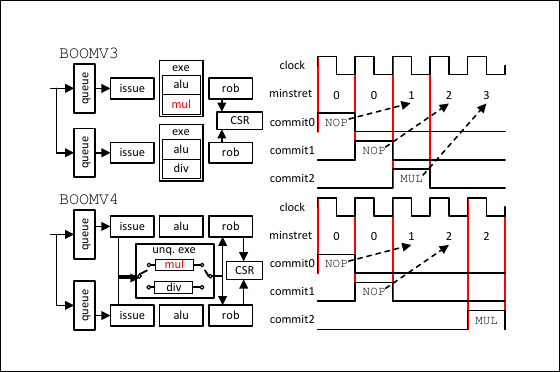}
    \caption{
    \boomt{} and \boomf{} have the same bug due to reusing modules that update the \texttt{minstret} register two cycles after an instruction is committed.
    \boomf{} has more variants of the bug due to its new microarchitectures.
    For example, the \texttt{MUL} instruction will trigger the bug in \boomf{} but not in \boomt{}.
    Red lines highlight the clock cycles when \texttt{minstret} is accessed to represent architectural states, while dashed arrows point to the actual commit number for each instruction.}
    \label{fig:case_study}
\end{figure}

\noindent\textbf{Shared root cause} lies in the interaction between the reorder buffer (\texttt{ROB}) and the \texttt{CSR} modules. 
The \texttt{ROB} module manages instruction commit logic, while the \texttt{CSR} module tracks architectural states, including privilege levels and the number of committed instructions (i.e., \texttt{minstret}).
In both \boomt{} and \boomf{}, committing an instruction by the \texttt{ROB} module triggers a two-cycle delay to update the \texttt{minstret} register in the \texttt{CSR} module, as shown by the waveforms in Figure~\ref{fig:case_study}.
However, the processors are configured to expose architectural states to the software level when an instruction is committed without counting this delay.
This makes the CSRs, such as \texttt{minstret}, inaccurately represent the processor's architecture states, leading to some instructions failing to increment the \texttt{minstret} register, while others increment the register by two,
\boomt{} and \boomf{} reuse both modules, thereby sharing the same bug.

\noindent\textbf{Variants in \boomf{}.} However, \boomf{} includes 10 additional instructions that can trigger this bug due to different microarchitectures.
Both \boomt{} and \boomf{} are superscalar processors, capable of executing multiple instructions in parallel via multiple issues~\cite{patterson2016computer}. 
In Figure~\ref{fig:case_study}, they are configured with two issue slots.
In \boomt{}, each issue slot is equipped with an execution unit, including a multiplier for the \texttt{MUL} instruction.
In contrast, \boomf{} uses a unique execution unit across issue slots.
To avoid data races, the shared unit implements a static arbiter to serialize access, introducing extra latency when \boomf{} \red{accessing}\blue{accesses} the multiplier.
Consequently, executing the same test (an instruction sequence with two \texttt{NOP} and one \texttt{MUL}) results in \texttt{MUL} incrementing \texttt{minstret} by two on \boomf{}, but by one on \boomt{}.
Figure~\ref{fig:case_study} illustrates this discrepancy, and Appendix~\ref{apx:bugs} lists all instructions observed to trigger this bug.
This case highlights that (i) the design reuse strategy can propagate vulnerabilities across processor generations, and (ii) directly reusing tests from prior processors may fail to detect variants of vulnerabilities. For example, \boomf{} has 10 additional variants of the vulnerability.
Properly mutating PP-tests (e.g., by altering instruction opcodes~\cite{kande2022thehuzz}) increases the probability of detecting such variants.


However, effective mutation is non-trivial. 
Over- or under-mutating a test can reduce a fuzzer’s efficiency and effectiveness. 
Fuzzers must carefully balance how long to mutate a given test versus when to proceed to the next.

\begin{tcolorbox}[colback=gray!10, colframe=black!50, boxrule=0.5pt, arc=0pt, left=2mm, right=2mm, top=1mm, bottom=1mm]
\textbf{Observation \#1:} 
Mutating PP-tests is essential for uncovering variants of vulnerabilities in PUTs. 
However, the effectiveness and efficiency of fuzzing depend on balancing how long to mutate a test before switching to the next one.
\end{tcolorbox}

\subsection{Can Reusing PP-tests Enhance Coverage?}~\label{sec:mot_cov}
Since improving coverage achieved by a fuzzer is critical for detecting vulnerabilities, we evaluate whether simply reusing PP-tests can improve coverage on PUTs.
\input{Table/story}

\noindent\textbf{Evaluation setup.} 
We evaluate coverage achieved by a fuzzer in terms of \textit{\tcov{}}\footnote{Total coverage refers to the cumulative percentage of coverage points reached after executing a given number of tests.}
and \textit{\spd{}}.
We use an existing fuzzer~\cite{kande2022thehuzz} to fuzz \boomt{} (the prior processor) and \boomf{} (the PUT)~\cite{boom} by generating 21K tests.
The 21K tests from \boomt{} serve as PP-tests.
We define \spd{} as the number of tests required to reach a given total coverage.
For example, the baseline fuzzer reaches $66\%$ \tcov{} after generating $8,548$ tests, whereas directly reusing tests from \boomt{} requires $11,246$ tests to reach the same \tcov{}, resulting in a $0.76\times$ slowdown. 
Also, executing all PP-tests from \boomt{} on \boomf{} resulted in $0.19\%$ less \tcov.
Table~\ref{tab:reuse_test} summarizes the coverage results of all strategies, with \ourtool{} performing the best (see Section~\ref{sec:implementation} for the details of the evaluation setup).

\noindent\textbf{Inadequate diversity.} 
Directly reusing tests from \boomt{} does not achieve higher \tcov{} than baseline on \boomf{} because the PP-tests cannot explore unique design features of \boomf{}.
To address the issue, we collect 21K tests each from two RISC-V processors \rc{}~\cite{rocket_chip_generator} and \cva{}~\cite{cva6} (totally 63K tests), which are widely used as benchmarks by processor fuzzers~\cite{hur2021difuzzrtl,chen2023hypfuzz,xu2023morfuzz,solt2024cascade,wugenhuzz}.
We randomly pick 21K tests from all three processors (i.e., Random tests from procs).
The strategy achieves $0.26\%$ less \tcov{} than the baseline.

In contrast, using the same PP-tests, \ourtool{} achieves $3.92\%$ more \tcov{} compared to the baseline. 
This highlights that simply reusing PP-tests achieves similar \tcov{} as fuzzing the PUT directly.
However, mutating PP-tests, as \ourtool{} does, can help explore unique design features of the PUT, leading to higher \tcov{}.

\begin{tcolorbox}[colback=gray!10, colframe=black!50, boxrule=0.5pt, arc=0pt, left=2mm, right=2mm, top=1mm, bottom=1mm]
\textbf{Observation \#2:} 
Simply reusing PP-tests achieves similar total coverage as fuzzing the PUT directly.
Mutating PP-tests achieves more coverage.
\end{tcolorbox}


\noindent\textbf{Inefficient order of execution.}
The naive approach for reusing PP-tests is to execute them on the PUT in the identical sequence (as they were executed during the PP fuzzing campaign).
However, this method proves insufficient for enhancing the \spd{}, as it degrades the \spd{} by a factor of $0.76\times$ compared to fuzzing the PUT from scratch, evidenced by our experimental results (see second row in Table~\ref{tab:reuse_test}). 
As expected, results from other PPs with this method demonstrate even worse performance.

A more advanced approach would be to randomly select tests from the PP-tests and execute them on the PUT. 
However, this method encounters the same inefficiency problem of \spd{}, as demonstrated by our experiments (see third row in Table~\ref{tab:reuse_test}). 
This approach, on average, further degrades the \spd{} by a factor of $0.74\times$.
This shows that the order of execution impacts the \spd{}.

To advance the method, we ranked the tests based on their \textit{\scov{}}\footnote{Standalone coverage refers to the percentage of coverage points achieved by executing a single test.}, while executing on their related PP. 
For instance, if a PP-test originated from \rc{}, we ranked it based on its \scov{} from \rc{}. In the fourth row of Table~\ref{tab:reuse_test}, we present results for the most closely related PP, \boomt{}, to our PUT, \boomf{}, which demonstrates that while ranking provides better \spd{} compared to identical sequence and random selection, it still delivers the same performance as the baseline. 
To further advance this method, we ranked the PP-tests based on their average \scov{} across all three processors, achieving a $2.30\times$ improvement in \spd{} (see fifth row in Table~\ref{tab:reuse_test}). 
These results suggest that leveraging a broader set of PPs helps identify highly effective tests.

While the aforementioned strategies improve \spd{}, their static nature prevents them from improving \tcov{}. This observation motivates\red{ the idea} that a dynamic strategy, while benefiting from effective test reuse and improving \spd{}, could also increase \tcov{}. 
We propose the\red{ design of the} \ourtool{} framework to address this lack of dynamic strategy. \ourtool{} improves the \spd{} by $5.40\times$, outperforming all evaluated strategies. 
The key insight is that a test's effectiveness in increasing coverage varies during the fuzzing process. 
For instance, a test achieves an average $4.99\%$ \textit{\icov{}}\footnote{Incremental coverage refers to the percentage of newly reached coverage points compared to \tcov{} by a test.} across all three processors when the \tcov{} is $55\%$, but the same test achieves only $0.06\%$ \icov{} when the \tcov{} is $70\%$. 

\begin{tcolorbox}[colback=gray!10, colframe=black!50, boxrule=0.5pt, arc=0pt, left=2mm, right=2mm, top=1mm, bottom=1mm]
\textbf{Observation \#3:} 
The effectiveness of tests varies during the fuzzing process.
The fuzzer needs to dynamically evaluate the effectiveness of tests and determine which test to mutate.
\end{tcolorbox}



%% file: Table/story.tex
\begin{table}
\caption{Coverage results of different test reuse strategies on \boomf{}~\cite{boom}. ``PP'' refers to \cva{}, \rc{}, and \boomt{}.}
\label{tab:reuse_test}
\centering
\resizebox{0.85\columnwidth}{!}{%
\begin{tabular}{ccc}
\hline
Strategy             & \begin{tabular}[c]{@{}c@{}}Ave. Total \\ Coverage ($\%$)\end{tabular} & Speedup \\ \hline
Fuzzing from Scratch (baseline)           & $66.66$                 & $1.00\times$    \\ \hline
Same Sequence of \boomt{}-Tests        & $66.47$                 & $0.76\times$    \\ \hline
Random Sequence of PP-Tests & $66.40$                 & $0.74\times$    \\ \hline
Ranked Sequence of \boomt{}-Tests & $66.64$                 & $1.00\times$    \\ \hline
Ranked Sequence of PP-Tests & $66.76$                 & $2.30\times$    \\ \hline
\ourtool{}     & $70.58$                 & $5.40\times$   \\ \hline
\end{tabular}%
}
\end{table}

%% file: draft/method.tex
\section{Methodology}\label{sec:method}

In this section, we first discuss why CB is suitable for reusing prior-processor test (PP-tests) with hardware fuzzing and how we model hardware fuzzing as a CB problem.
We then describe how we build and train the CB model. 
We improve the efficiency of the CB model based on the characteristics of fuzzing (we call it adaptive CB), and we further optimize the training tests to enhance the effectiveness of the CB model.
Finally, we integrate the trained CB model with two distinct fuzzers to test its effectiveness, demonstrating that the approach is agnostic to any hardware fuzzers.



\subsection{Why Contextual Bandit (CB)?}~\label{sec:why_cb}
As shown in~\ref{sec:motivation}, a naive, greedy approach that prioritizes historically high-performing tests risks converging on local optima. 
Conversely, an exhaustive evaluation of all prior test cases is computationally prohibitive and lacks the dynamic and adaptive nature. 
This problem of choosing an action (a test) based on the current state (\textit{\ccontext{}}\footnote{Coverage context refers to the point in the fuzzing process when total coverage reaches specific thresholds, such as $55\%$, $60\%$, or $65\%$.}) to maximize cumulative reward (\tcov{}) is precisely a \textbf{contextual decision-making problem}. 

The CB algorithms are suited to this challenge due to three key properties.
First, CBs incorporate \textit{context} into their policy. 
The effectiveness of a test changes as fuzzing progresses; tests for broad exploration at low \tcov{} (e.g., $50\%$) are different from those needed for deep, corner-case exploration at high \tcov{} (e.g., $70\%$). 
CBs learn a policy that adapts to the evolving coverage context, enabling \textbf{adaptive decision-making} for \ourtool{} to select the most effective tests at any coverage context. 
Second, CBs are designed to balance leveraging known effective tests (exploitation) with investigating new ones (exploration). 
This balance is critical for fuzzing efficiency. \ourtool{} uses this capability to dynamically adjust its strategy (\textbf{exploration-exploitation trade-off}), ensuring it efficiently uses highly effective tests. 
Third, CBs are computationally lightweight and can handle a large action space list PP-test corpora without the overhead of more complex reinforcement learning models. Their \textit{anytime learning}~\cite{lattimore2020bandit} property means they continuously refine their policy and retain a useful solution even if the training process is interrupted (\textbf{scalability and anytime learning}).

\subsection{Modeling Fuzzing as a Contextual Bandit (CB) Problem}~\label{sec:model_cb}
To apply CB to fuzzing, we model the interaction between \ourtool{} and the PUT as a contextual bandit problem.
During training, \ourtool{} learns a CB policy by iteratively exploring PP-tests and observing their effectiveness.
The goal is to identify and prioritize tests that maximize \tcov{}.

\noindent\textbf{Preliminary formulation.}
The primary components of a bandit problem are: \textit{agent}, \textit{environment}, \textit{set of arms}, \textit{context}, \textit{policy}, and \textit{reward}, as mentioned in Section~\ref{sec:bg_bandit}.
\ourtool{} is the agent that learns the policy (i.e., which test to select first) through interacting with the environment. 
We let the training environment be the PUT and its verification environment, such as software simulators~\cite{vcs}, which provides coverage feedback.
We then let $n$ be the total number of training steps.

\begin{defn}\label{defn_1}
   $\mathcal{A}$ is the finite set of \textbf{arms}. 
   It contains all PP-tests. 
   $a_t$ denotes the test selected by \ourtool{} at time $t$.
\end{defn}

\begin{defn}\label{defn_2}
    $\mathcal{C}\subseteq[0,1]$ is the finite set of \textbf{\ccontext{}} which contains all possible \tcov{} of a PUT achievable by the fuzzer.
    We use $c_t\in\mathcal{C}$ to denote the \ccontext{} at time $t$, where $c_t$ ranges from $0$ (covers no point) to $1$ (covers 100\% coverage points).
\end{defn}

\begin{defn}\label{defn_3}
    $r_t(c_t, a_t)$ denotes the reward revealed by \ourtool{} at time $t$ after executing the test $a_t$ selected under context $c_t$. 
    To maximize \tcov{}, we let the reward represent \textbf{\covi{}} after receiving coverage feedback for the selected test from the environment as $r_t(c_t, a_t) = \Delta cov_{t}(a_t)$, where $\Delta cov_{t}(a_t) = \{\Delta cov_{t}\mid \Delta cov_{t} \text{ is covered by } a_t \text{ at } t \text{ but not } c_t$\}. 
    In the training stage, the \covi{} is the \icov{} of the selected test $a_t$.
\end{defn}

At each time step, \ourtool{} observes the current \ccontext{}. It chooses a test, runs the test, and checks how much \covi{} is achieved (reward). It then updates its policy about which tests are most effective at increasing coverage under which \ccontext{}.
Thus, the goal is to learn the optimal CB policy that maximizes the sum of collected rewards, i.e., \tcov{}. 
Let $g\colon\mathcal{C}\times\mathcal{A}\to\mathcal{C}$ denote the deterministic function that represents the PUT and its verification environment. 
In particular, given the coverage context of a PUT $c_t$ and the test $a_t$ selected by \ourtool{}, $c_{t+1}=g(c_t,a_t)$ is the next coverage context of the PUT. Formally, the mathematical objective of \ourtool{} is 
\begin{equation*}
    \max_{\pi} \mathbb{E}\bigg[\sum_{t=1}^n r_t(c_t, a_t) \mid a_t\sim\pi(\cdot\mid c_t), c_{t+1} = g(c_t,a_t) \bigg].
\end{equation*}

\subsection{Training a Contextual Bandit (CB) Model}\label{sec:train_cb}
\begin{figure}[!th]
    \centering
    \includegraphics[width=0.95\linewidth,trim={14 18 20 20}, clip]{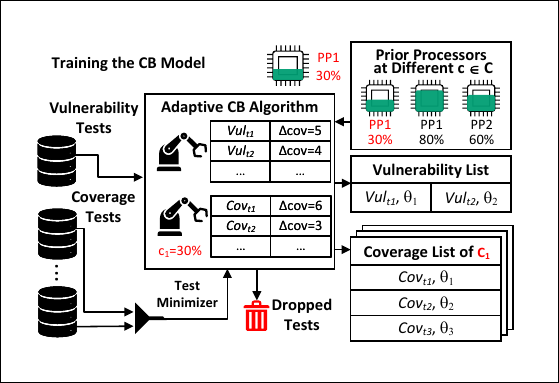}
    \caption{\ourtool{}'s training stage. $C$ is the set of different coverage contexts. $Vul_t$ is a test in the vulnerability list, and $Cov_t$ is a test in the coverage list.}
    \label{fig:learn_stage}
\end{figure}

Figure~\ref{fig:learn_stage} shows how \ourtool{} applies CB to identify effective tests\red{. The training stage} \blue{and} takes two inputs: PP-tests and various \ccontext{}s collected during fuzzing PPs.
Following industry practices in creating directed tests for different verification purposes~\cite{seedreuse}, we categorize PP-tests into either \textbf{vulnerability tests} that trigger known vulnerabilities\red{ in one or multiple PPs} or \textbf{coverage tests} that improve coverage achieved on one or multiple PPs.

During training, \ourtool{} generates two types of \textit{test lists}: \textit{\vlist{}} and multiple \textit{\clist{}s}.
 A test list contains optimal PP-tests identified by the CB model associated with a probability distribution, in which $\theta_1$ corresponds to the probability of Test1 being selected by the CB model for a given coverage context $c_1$.
Each \clist{} is tuned to a specific \ccontext{}. 
Multiple coverage lists help \ourtool{} to prioritize tests precisely based on the current \ccontext{} and prevent \ourtool{} from selecting the same test across different \ccontext{}s.
For example, the \clist{} trained for $50\%$ \ccontext{} contains more tests that explore major design spaces in a PUT. 
While the \clist{} for $70\%$ \ccontext{} contains more tests that explore corner cases (see Section~\ref{sec:anal_dist}). 
The number of \clist{} depends on the granularity of \ccontext{} configured during \ourtool{}'s setup. 

However, using PP-tests to train a CB model is non-trivial due to the limitations in the original CB algorithms.
To address this, we first use a \textit{test minimizer} to preprocess PP-Tests and remove redundant ones that reach the same coverage points. 
The minimized tests are then used by the \textit{adaptive CB algorithm}, which drops ineffective tests and learns which tests are most effective at achieving coverage increment under different coverage contexts. 

\begin{figure}[h]
    \centering
    \includegraphics[width=0.85\linewidth,trim={7 4 8 6}, clip]{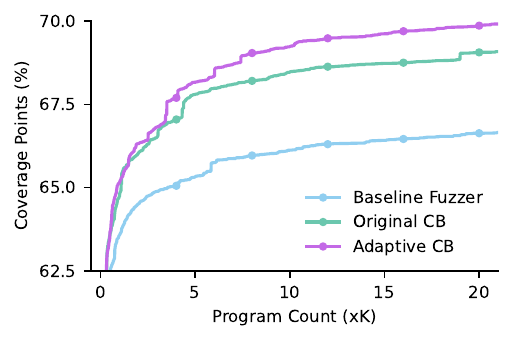}
    \caption{The \tcov{} achieved by the baseline fuzzer, original CB, and adaptive CB on \boomf{}~\cite{boom}.}
    \label{fig:cov_cbs}
\end{figure}

Figure~\ref{fig:cov_cbs} shows the coverage achievement of the original CB algorithm, the adaptive CB algorithm, and the baseline fuzzer. 
The original CB algorithm achieves $2.43\%$ more coverage and $4.35\times$ speedup than the baseline. 
The adaptive CB algorithm achieves $3.92\%$ more coverage and $5.40\times$ speedup\footnote{To make a fair comparison, the results only use PP-tests from the same baseline fuzzer. We collect PP-tests from multiple baseline fuzzers, which achieves substantial improvement (see Section~\ref{sec:cov_eva}).}.

\subsection{Challenges and Solutions}\label{sec:challenge_sol}
The original CB algorithm achieves limited improvement on the baseline fuzzer due to two major challenges.
\noindent\textbf{Challenge 1.} 
Ineffective tests remain in the lists.
The original CB algorithm assumes that all actions remain available forever~\cite{lattimore2020bandit}. 
However, in fuzzing, some tests consistently achieve zero coverage increment and should not remain in consideration. 
Moreover, since training a CB model is an iterative process, newly selected tests may outperform existing ones by achieving higher coverage increments. 
Without a mechanism to eliminate ineffective tests, they remain in the test list, degrading \ourtool{}'s overall performance.

\noindent\textbf{Solution 1.} We develop an \textit{adaptive CB algorithm} with an \textit{elimination function}.
Unlike conventional elimination strategies~\cite {lattimore2020bandit} that only eliminate suboptimal arms, our approach also aims to explore a broad range of effective tests to enhance fuzzing efficiency. 
The algorithm trains the model by iteratively identifying and retaining highly effective tests under different coverage contexts. 
It monitors their effectiveness through moving-average coverage increments and dynamically eliminates ineffective ones.
Tests that consistently achieve high coverage increments are promoted to the final model.


Algorithm~\ref{algo:adapt_cb} outlines the adaptive CB algorithm, with the elimination function highlighted in gray.
\blue{
The inputs are the PP-test corpus $\mathcal{A}_{\text{corpus}}$, a \ccontext{} $c$, the number of arms $k$, the check window $\gamma$, the pre-defined adaptive threshold $\theta$, and the training step $n$.
The outputs are the coverage list $\mathcal{A}$ and the policy $\pi$.}
Since the number of tests (arms) is changing during training, the algorithm uses an auxiliary policy $\pi_{\mathrm{tmp}}$ and a temporary arm set $\mathcal{A}_{\mathrm{tmp}}$ that always contains $k$ arms.
For each coverage context $c$, the algorithm calculates the average coverage increments a test can achieve (Line \red{6}\blue{8}).

Once the test has been selected at least $\gamma$ times (Line \red{7}\blue{9}), the algorithm evaluates its effectiveness.
If the test achieves no coverage increment, it is dropped from $\mathcal{A}_{\mathrm{tmp}}$ with its relevant variables (Line \red{9}\blue{11}), and a new test from the PP-test corpus $\mathcal{A}_{\mathrm{corpus}}$ is added to maintain the number of arms (Lines \red{10-11}\blue{12--13}).
However, if the probability to select the test exceeds the pre-defined adaptive threshold $\theta$ (Line \red{12}\blue{14}), it is promoted to the ultimate arm set $\mathcal{A}$ (i.e., the test list) and the final policy $\pi$ (Lines \red{13-14}\blue{15--16}). 
Since the number of the ultimate arms will vary depending on the coverage context and the effectiveness of PP-tests, the final policy needs to be normalized to ensure it remains a valid probability distribution (Line \red{18}\blue{20}).

\input{Code/algo1}

\noindent\textbf{Challenge 2.} Tests may achieve the same coverage points.
The CB algorithm evaluates the effectiveness of tests independently, without accounting for overlap in coverage. 
As a result, it may prioritize multiple tests that reach the same coverage points, even if each individually achieves a high coverage increment. 
This redundancy reduces exploration diversity and leads to inefficient use of fuzzing resources.

\noindent\textbf{Solution 2.} We develop the \textit{test minimizer} to remove redundant tests before training, ensuring computational efficiency while preserving the \tcov{} achieved by the original test set.
Specifically, the test minimizer selects the smallest possible subset of PP-tests that together reach the same total coverage as all PP-tests. 
For example, if test $a$ covers all the coverage points covered by tests $b$ and $c$, then tests $b$ and $c$ are removed. 

However, identifying such redundancies is computationally challenging.
Comparing tests pairwise or exhaustively evaluating all subsets quickly becomes intractable, as the number of combinations grows exponentially with the number of tests.
Therefore, inspired by \mints{}~\cite{hsu2009mints}, we formulate the task of selecting a minimal subset of tests as an optimization problem and solve it using integer programming. 
The model of test minimizer is discussed in Appendix~\ref{apx:appendix_model}.
The model requires a \textbf{coverage matrix} as the input, where each row corresponds to a single test and shows which coverage points it reaches. 
The number of columns represents the total number of coverage points defined by a target coverage metric in a PUT.

\noindent\textbf{Constructing the coverage matrix} for our test minimizer involves two key requirements. First, we must understand the \textbf{hierarchical structure} of the PUT. 
This is essential for correctly attributing coverage points to their corresponding register-transfer level (RTL) modules, enabling fine-grained analysis of where each test explores within the design. 
Second, we need to monitor which coverage points within each RTL module are reached by individual tests. 
This enables us to compile a \textbf{row vector} for each test in the coverage matrix, where each entry reflects whether a coverage point was covered.


To address these requirements, we leverage the coverage databases generated by \textit{Synopsys} \vcs{}~\cite{vcs}, an industry-standard simulator.
While \vcs{} does not provide APIs for directly accessing the hierarchical structure of the PUT or the status of coverage points, we can extract this information by parsing the internal files of the databases. We then use the information to concatenate row vectors for tests and construct the coverage matrix.
The approach ensures stability and remains compatible with a broad range of processor designs.
As a result, it facilitates the training of our CB model and\red{ its} \blue{\ourtool{}'s} integration with\red{ existing} industrial verification flows. The result shows that test minimizer successfully identifies the minimal subset by removing \avereduction{} redundant tests, reducing the number of tests from $126K$ to\red{ around} $1.5K$ (See Section~\ref{sec:anal_dist}).

\noindent\blue{\textbf{No elimination is performed on vulnerability tests.}} Note that the elimination function and the test minimizer are applied only to coverage tests, which are\red{ typically} numerous and highly redundant. 
In contrast, we assume that vulnerability tests trigger distinct vulnerabilities and explore diverse design spaces. Therefore, no vulnerability tests are dropped during training.

\subsection{Integrating the Trained CB Model with Processor Fuzzers}\label{sec:integrate_cb}
\noindent\textbf{Environment for testing.} To evaluate the generalizability of our CB model, we test it in an environment that differs from the training stage.
Given that fuzzers may employ different mutation strategies~\cite{kande2022thehuzz,solt2024cascade,xu2023morfuzz}, the testing \textit{environment} includes both the baseline fuzzer and the PUT. 
Unlike the training stage, where the model uses only the \icov{} of each PP-test as the reward, \ourtool{} evaluates the cumulative \icov{} of the PP-test and its mutated variants. 
Based on the cumulated \icov{}, the CB model automatically decides whether to continue mutating the current test or proceed to the next one.

Altering the testing environment is standard practice for assessing the generalizability of a CB model.
This is analogous to sim-to-real generalization in reinforcement learning~\cite{panaganti-rfqi}, where models trained in simulation are tested under altered dynamics, such as changes in gravity or actuator noise. 
Similarly, testing \ourtool{} in an environment with different fuzzing mechanisms provides a more robust assessment of its effectiveness.
Empirical results show that \ourtool{} is agnostic to fuzzers with distinct mechanisms and outperforms them in both \tcov{} and \spd{} (see Section~\ref{sec:cov_eva}).

\noindent\textbf{Resuming seed generation.} 
Unlike the training stage, \ourtool{} does not add new tests to the curated test lists during testing. 
Thus, when a coverage list becomes empty due to the unique design features of the PUT, \ourtool{} allows the fuzzer to resume its native seed generation to continue exploring the design spaces. 
Additionally, if the \tcov{} is too low (no match with any \ccontext{}) for the CB algorithm to select tests from the next \clist{}, and the current coverage list is empty, \ourtool{} will skip directly to the next list, avoiding stalling fuzzing progress. 
In summary, \ourtool{} resumes the fuzzer’s own seed generation strategies only after all curated test lists have been exhausted, ensuring efficient reuse of PP-tests while maintaining flexibility to explore unique design features of the PUT.


\ourtool{} begins by selecting tests from the vulnerability list and monitors their coverage increment. 
If a test and its mutated variants fail to improve coverage, it is dropped. 
Once all tests in the vulnerability list are dropped, \ourtool{} switches to coverage lists.
\ourtool{} tracks the total coverage achieved by vulnerability tests and uses this information to determine the current coverage context and decide the starting coverage list.
\ourtool{} periodically evaluates each test, removing those that show no further coverage increment, and continues this process until all coverage lists are empty. Finally, \ourtool{} resumes the seed generation technique of the processor fuzzer to explore the unique design features of the PUT.
Appendix~\ref{apd:integrate_cb} shows the detailed integration process.

\subsection{Putting it all Together}\label{sec:put_together}

\begin{figure}[h]
    \centering
    \includegraphics[width=0.95\linewidth, trim={21 16 21 18},clip]{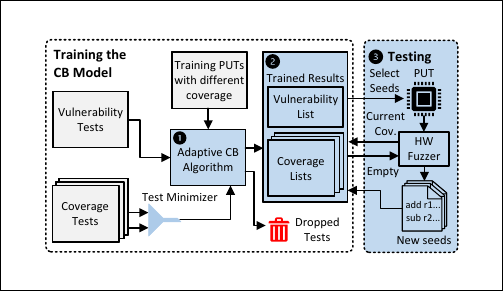}
    \caption{The framework of \ourtool{}.}
    \label{fig:framework}
\end{figure}

\red{The \ourtool{} framework consists of two main stages: \textbf{training} and \textbf{testing}, as shown in Figure~\ref{fig:framework}.}
\blue{The \ourtool{} framework consists of the \textbf{training} and the \textbf{testing} stages, as shown in Figure~\ref{fig:framework}.}
In the training stage, \ourtool{} begins by preprocessing a large corpus of tests from prior fuzzing campaigns.
To address\red{ redundancy of test} \blue{test redundancy}, a test minimizer removes \avereduction{} of redundant tests, resulting in a lean and effective test corpus for training.
Next, \protect\circleicon{1} the adaptive CB algorithm analyzes the performance of\red{ these} minimized tests across prior processors and \ccontext{}s. 
This allows \ourtool{} to learn which tests are most likely to increase coverage \red{in}\blue{under} specific \ccontext{}s.
\protect\circleicon{2} The output is a set of carefully curated test lists: a vulnerability list for known vulnerabilities, and several coverage lists, each tuned to a\red{ specific} \ccontext{}.
In \protect\circleicon{3} the testing stage, \ourtool{} integrates with an existing processor fuzzer to guide its test selection. 
\red{The fuzzer starts with the pre-trained vulnerability list to identify variants of known vulnerabilities. 
As the fuzzer runs, \ourtool{} monitors the \tcov{} and dynamically selects the most appropriate coverage list based on the current \ccontext{}. 
This guided approach ensures that the fuzzer always has access to highly effective seeds, maximizing its efficiency.
When all curated lists are exhausted, \ourtool{} resumes the fuzzer's naive seed generation to continue exploring the unique design features of the PUT.} 
The result is a unified framework that combines the speed and targeted nature of directed testing with the explorative power of processor fuzzers.

\section{Benchmarks, Training, and Implementation}~\label{sec:implementation}
\noindent\textbf{Fuzzer selection.} 
We select fuzzers to generate tests for training based on two criteria: 
(i)~Do fuzzers implement distinct algorithms that increase the probability of verifying diverse processor functionalities? 
This diversity helps \ourtool{} identify tests that explore a broad range of design spaces.
(ii)~Are fuzzers' strategies generalizable, allowing them to be applied across different processor designs? 
This generalizability enables \ourtool{} to evaluate the effectiveness of a test on a variety of processors following the same ISA.
Based on these criteria, we select \thehuzz{} and \cascade{}~\cite{solt2024cascade} as our baseline fuzzers. Appendix~\ref{apd:fuzz} includes detailed justifications.
\red{The training results show that \cascade{}'s tests are particularly effective at exploring commonly executed design spaces, while \thehuzz{}'s tests tend to explore corner cases.
These complementary strengths enhance the diversity of PP-tests and improve \ourtool{}'s ability to explore a wide range of design spaces in PUTs.}
\red{Appendix~\ref{apd:testdist} analyzes the distributions of PP-tests from both fuzzers in detail.}

\noindent\textbf{Benchmark selection.} 
While ISAs define the functionalities and interfaces of processors, hardware vulnerabilities often arise from differences in microarchitectures.
Therefore, we select processors with diverse microarchitectures as benchmarks. 

\input{Table/benchmark_info}

Given that most commercial processors are protected intellectual property~(IP) and closed-source, we select benchmarks from open-source \riscv{} processors.
Table~\ref{tab:bench} summarizes the microarchitectures of each processor. 
To capture a broad range of microarchitectures, we select three widely-used \riscv{} processors \blue{by existing processor fuzzers~\cite{kande2022thehuzz,chen2023hypfuzz,solt2024cascade,xu2023morfuzz,wugenhuzz}} for training: \rc{}~\cite{rocket_chip_generator}, \boomt{}~\cite{boom}, and \cva{}~\cite{cva6}.
\red{Compared to \rc{}, \cva{} and \boomt{} include more complex microarchitectures such as out-of-order execution~(OoO) and single instruction-multiple data~(SIMD).}
We also include \boomf{} and \rsd{}~\cite{mashimo2019open} to evaluate the effectiveness of \ourtool{}.
\boomf{} is the next generation of \boomt{}, enabling assessment of \ourtool{}'s adaptability to different generations of processors.

\noindent\textbf{Simulation and vulnerability detection.} We use \textit{Chipyard} (version 1.13.0, commit \texttt{69eba86})~\cite{chipyard} as the simulation environment for \cva{}~\cite{cva6}, \rc{}~\cite{rocket_chip_generator}, \boomt{}, and \boomf{}~\cite{boom}, while \rsd{} (commit \texttt{7b65f6b})~\cite{mashimo2019open} is simulated using \vcs{}~\cite{vcs}.
We employ \textit{differential testing}, a commonly used approach in
the processor fuzzers, to detect vulnerabilities and bugs, as mentioned in Section~\ref{sec:bg_hf}.
We use \texttt{Spike} (commit \texttt{de5094a})~\cite{spike} as the golden-reference model, widely used by existing fuzzers~\cite{hur2021difuzzrtl,kande2022thehuzz,chen2023hypfuzz,gohil2024mabfuzz}. 
All experiments are conducted on a 48-core AMD EPYC 7443 processor at 2.6 GHz with 256GB RAM. 

\noindent\textbf{Building the PP-test corpus.}
For vulnerability tests, we collect tests that trigger vulnerabilities detected by two fuzzers on the training processors. 
\thehuzz{} detects four vulnerabilities in \cva{}, while \cascade{} detects ten vulnerabilities in \cva{} and two in \boomt{}.
For coverage tests, we run both \thehuzz{} and \cascade{} to generate 21K tests per training processor, resulting in a total of 126K tests.
For a fair comparison in Section~\ref{sec:motivation}, we use \thehuzz{}~\cite{kande2022thehuzz} and its coverage tests as the baseline.


\subsection{Training}\label{sec:train}
\noindent\textbf{Preprocessing the PP-test corpus.}
\red{We implement its optimization model using DOcplex~\cite{ibm_cplex}, provided by IBM.}
As mentioned in Section~\ref{sec:train_cb}, we develop a test minimizer to remove redundant tests \blue{and implement it using DOcplex~\cite{ibm_cplex} from IBM.}
To construct the coverage matrix, we developed a\red{ custom} parser \red{that}\blue{to} \blue{parse}\red{processes}\red{ the internal files of} \vcs{} coverage databases (i.e., \texttt{.vdb}).
After executing each test, the parser identifies the RTL modules and their associated coverage points, determines whether each point was reached, and encodes this as a binary vector, with the ordering aligned to the traversal sequence of the module hierarchy. 
This binary vector serves as a row in the coverage matrix. 
The parser repeats this process for all PP-tests.


\noindent\textbf{Sampling coverage contexts.}
We sample coverage contexts from the training processors during fuzzing with \thehuzz{} and \cascade{}. 
We obtain coverage feedback using \textit{Synopsys} \texttt{VCS}~\cite{vcs}\red{, a standard industry simulation tool}. 
Starting at $55\%$ total coverage, we sample a new coverage context at every $5\%$ increment\red{ in coverage}. 
\red{The sampling frequency depends on the coverage metrics as discussed below.}

\noindent\textbf{Training stage.}
We implement both the original and adaptive CB algorithms using the \texttt{MABWiser} Python library~\cite{StrongKK21}.
To encourage exploration of diverse, high-effective tests, we adopt the $\epsilon$-greedy algorithm for the CB algorithm with an exploration probability ($\epsilon$) of 0.2.
We use branch coverage as the target metric due to its importance in uncovering vulnerabilities~\cite{chen2023hypfuzz}. 
\ourtool{} can also be configured for other coverage metrics.
We observe that branch coverage often starts above $50\%$ and can quickly exceed $60\%$, but growth slows around $65\%$ across the three training processors. 
Based on this observation, we define \ccontext{} at $\{55\%, 60\%, 65\%, 70\%\}$.
These coverage contexts are also applied during testing to determine appropriate coverage lists.

To maximize effectiveness and avoid test overlap between coverage lists, we begin training \ourtool{} from the highest \ccontext{} (\blue{e.g.,} $70\%$) and use the remaining tests to train the lower contexts. 
During training, \ourtool{} randomly selects one of the training processors to compute the reward of each test for a given coverage context.

\noindent\textbf{Configuring the CB model.} According to Algorithm~\ref{algo:adapt_cb}, we heuristically configure the number of arms $k$ to \trainarm{} and check window $\gamma$ to \trainwindow{}.
\red{The configuration ensures that each coverage list can maximally contain \trainarm{} tests.}
\ourtool{} is trained at \trainhorizon{} steps to identify effective tests for each coverage context.
A critical parameter in the adaptive CB algorithm is the adaptive threshold $\theta$, which determines \red{the effectiveness of tests}\blue{test effectiveness}.
Setting $\theta$ too high may result in too few tests being identified (fewer than 10), while setting it too low may include too many, diluting the selection of optimal tests.
To balance this tradeoff, we configure $\theta$ separately for each coverage context: $\theta = \{55\%: 1.90, 60\%: 1.50, 65\%: 0.90, 70\%: 1.26\}$.
The \red{configuration}\blue{threshold} ensures that, by the end of training, each coverage list includes approximately \trainarm{} effective tests, aligning with the number of arms $k$. \blue{Appendix~\ref{apx:fine_tune_threshold} details how we fine-tune the threshold automatically.}


\noindent\blue{\textbf{Manual efforts and automation.}} 
\blue{
Manual efforts are required to set up existing fuzzers for processors and collect coverage results because fuzzers may not support all benchmarks. This is a general step for all fuzzers. 
For vulnerability tests, we obtain them directly from artifacts or documented GitHub issues. 
Once the PP-test corpora and fuzzers are configured, the training and testing processes are fully automatic.
}

%% file: Code/algo1.tex
\begin{algorithm}[!h]
    \caption{Adaptive CB algorithm.}~\label{algo:adapt_cb}	
    \begin{algorithmic}[1]    
    \State \blue{\textbf{Inputs: $\mathcal{A}_{\text{corpus}}, c, k, \gamma, \theta, n$}}
    \State \blue{\textbf{Outputs: $\mathcal{A},\pi$}}
    \State \textbf{Initialize:} $\mathcal{A} \gets \emptyset$; $\mathcal{A}_{\text{tmp}} \subset \mathcal{A}_{\text{corpus}}$ with $|\mathcal{A}_{\text{tmp}}| = k$; $\pi_{\text{tmp}}(a | c) = \frac{1}{|\mathcal{A}_{\text{tmp}}|} \quad \forall a \in \mathcal{A}_{\text{tmp}}$;  $\pi(a | c) \gets 0 \quad \forall a$; $\hat{r}(a) \gets 0 \quad \forall a \in \mathcal{A}_{\text{tmp}}$; $\#(a) \gets 0 \quad \forall a \in \mathcal{A}_{\text{tmp}}$
        \State $\forall a \in \mathcal{A}_{\text{tmp}}$: update $\hat{r}(a)$ and $\#(a)$ once.
        \For{$t=1,2,\dots, n$}
        \State $a_t \sim \pi_{\text{tmp}}(\cdot \mid c)$; $r_t \gets r(c, a_t)$
        \State $\pi_{\text{tmp}} \gets \text{UpdatePolicy}(\pi_{\text{tmp}}, c, a_t, r_t)$
    \tikzmk{A}
        \State $\hat{r}(a_t) \gets \frac{\hat{r}(a_t)\times \#(a_t)+r(c,a_t)}{\#(a_t)+1}$; $\#(a_t) \gets \#(a_t) \mathrel{+} 1$
            \If{$\#(a_t) \geq \gamma$}
                \If{$\hat{r}(a_t) = 0$}
                    \State $\mathcal{A}_{\text{tmp}} \gets \mathcal{A}_{\text{tmp}} \setminus \{a_t\}$
                    \State $a' \gets \text{random\_sample}(\mathcal{A}_{\text{corpus}})$
                    \State $\mathcal{A}_{\text{tmp}} \gets \mathcal{A}_{\text{tmp}} \cup \{a'\}$
                \ElsIf{$\pi_{\mathrm{tmp}}(a_t\mid c) \geq \theta$}
                    \State $\mathcal{A} \gets \mathcal{A} \cup \{a_t\}$
                    \State $\pi(a_t\mid c)\gets\pi_{\mathrm{tmp}}(a_t\mid c)$
                    \State $\mathcal{A}_{\text{tmp}} \gets \mathcal{A}_{\text{tmp}} \setminus \{a_t\}$
                    \State $a' \gets \text{random\_sample}(\mathcal{A}_{\text{corpus}})$
                    \State $\mathcal{A}_{\text{tmp}} \gets \mathcal{A}_{\text{tmp}} \cup \{a'\}$
                \EndIf
            \EndIf
        \tikzmk{B}
        \boxitonept{default_gray}
        \vspace{-.2\baselineskip}
        \EndFor
        
        \State $\pi(a \mid c) \gets \frac{\pi(a \mid c)}{\sum_{a' \in \mathcal{A}}\pi(a' \mid c)}$\blue{; \textbf{return} $\mathcal{A},\pi$}
    \end{algorithmic}
\end{algorithm}

%% file: Table/benchmark_info.tex
\begin{table}[h]
\caption{Microarchitectural features in benchmark processors.\red{ \cva{}, \rc{} \, and \boomt{} are used for training. \boomf{} and \rsd{} are used for testing.}}
\label{tab:bench}
\resizebox{\linewidth}{!}{%
\begin{tabular}{|c|c|c|c|c|c|}
\hline
\textbf{Processor} & \textbf{OoO} & \textbf{SIMD} & \textbf{\begin{tabular}[c]{@{}c@{}}Ld/St\\ Forwarding\end{tabular}} & \textbf{\begin{tabular}[c]{@{}c@{}}Speculative\\ Scheduling\end{tabular}} & \textbf{\begin{tabular}[c]{@{}c@{}}Mem. Dep.\\ Predictor\end{tabular}} \\ \hline
\cva{}~\cite{cva6}               &       \tikzxmark{}       &      \tikzcmark{}         &    \tikzxmark{}                                                                 &      \tikzxmark{}   &    \tikzxmark{}              \\ \hline
\rc{}~\cite{rocket_chip_generator}                 &       \tikzxmark{}       &       \tikzxmark{}        &       \tikzxmark{}                                                              &                               \tikzxmark{}                                            &  \tikzxmark{}                                                                      \\ \hline
\boomt{}~\cite{boom}             &        \tikzcmark{}      &        \tikzxmark{}       &              \tikzcmark{}                                                       &                         \tikzxmark{}                                                  &     \tikzxmark{}                                                                   \\ \hline
\boomf{}~\cite{boom}             &        \tikzcmark{}      &        \tikzxmark{}       &              \tikzcmark{}                                                       &                         \tikzxmark{}                                                  &     \tikzxmark{}                                                                   \\ \hline
\rsd{}~\cite{mashimo2019open}               &        \tikzcmark{}      &       \tikzxmark{}        &   \tikzcmark{}                                                                  &               \tikzcmark{}                                & \tikzcmark{}                                                                       \\ \hline
\end{tabular}%
}
\end{table}

%% file: draft/evaluation.tex
\section{Evaluation}\label{sec:eva}
We evaluate \ourtool{} comprehensively to answer the following questions:
\begin{enumerate}
    \item[\textbf{Q1.}] Can \ourtool{} effectively reuse prior-processor tests (PP-tests) to detect variants of known vulnerabilities on processor-under-tests (PUTs)?
    \item[\textbf{Q2.}] Is \ourtool{} agnostic to existing fuzzers, and can it outperform them on \tcov{} and \spd{}?
    \item[\textbf{Q3.}] What are the individual contributions of various \ourtool{} optimizations and parameters?
\end{enumerate}

\red{In this section, }We first discuss the new vulnerabilities and bugs detected by \ourtool{}, highlighting their connection to the design reuse strategy. 
To show the generalization of \ourtool{} across different\red{ processor} fuzzers, we integrate \ourtool{} with \thehuzz{}~\cite{kande2022thehuzz} and \cascade{}~\cite{solt2024cascade} and evaluate the improvement on \tcov{} and \spd{}. 
Each fuzzer generates 21K tests per processor, and we repeat the experiment three times to evaluate the average results.
Finally, we analyze how the optimization and configuration impact the efficiency of \ourtool{}.

\subsection{New Vulnerabilities and Bugs}\label{sec:new_vul_bug}
\ourtool{} detected three new security vulnerabilities and two new functional bugs across multiple processors. \blue{Affected commits are mentioned in Section~\ref{sec:implementation}.}
\blue{\ref{b1} and \ref{b2} lead to incorrect outputs of performance counters.}
Bug~\ref{b1} exists in \rc{}~\cite{rocket_chip_generator}, \boomt{}, and \boomf{}~\cite{boom}, all of which share a\red{ common} hardware module. 
The bug exists across three processors due to \textbf{design reuse} and characteristics of the \textit{Chisel} hardware programming language~\cite{bachrach2012chisel}. 
\red{Specifically, \rc{} has a module called \texttt{CSR}, responsible for updating the values of control and status registers (CSRs), including \texttt{minstret}. 
\boomt{} and \boomf{} use the same module, thereby propagating the same bug. This bug also shows that the widely applied IP reuse strategy in hardware design can potentially propagate vulnerabilities to multiple processors.}
Bug~\ref{b2} affects both \boomt{} and \boomf{}, with \boomf{} exhibiting additional variants due to its new microarchitectures, as discussed in Section~\ref{sec:motivation}. 
\red{Though three processors use the same CSR module, they can include variants of bugs due to different microarchitectures.}
\blue{ 
Precise implementation of performance counters is critical for performance analysis, profiling, and debugging~\cite{demme_bottlenecks}. 
In addition, performance counters are used for malware detection~\cite{wang_performance_evaluation}. 
Therefore, instructions that produce incorrect counter values can reduce detection accuracy and may also serve as side channels for information leakage.}
A detailed analysis of these two bugs is provided in Appendix~\ref{apx:bugs}.

The vulnerabilities are severe, with CVSS scores ranging from $7.1$ to $8.5$ (out of $10$)~\cite{cvss_system}, and enable exploit scenarios, such as denial of service (DoS) or unauthorized access. 
The vendors either acknowledge the new vulnerabilities and bugs, or we have identified their root causes.
\red{We are applying Japanese Vulnerability Notes (JVNs) for the vulnerabilities \cite{jvn_website}.}
To answer \textbf{Q1}, \ourtool{} effectively reuses PP-tests to detect variants of known vulnerabilities on PUTs.
\noindent\begin{vul_ns}[\label{v1}]
\noindent \textbf{(CVSS score: 7.1).} 
A memory deadlock occurs on the \rsd{} processor when executing the \texttt{FENCE.I} instruction.
\blue{\ourtool{} reuses the vulnerability test that triggers a \texttt{FENCE.I}-based vulnerability in \cva{}~\cite{cva6} to detect this vulnerability.}
\blue{\ref{v1} presents a denial-of-service vector. Any user-mode program capable of executing arbitrary instructions can invoke \texttt{FENCE.I} to cause the memory deadlock of the processor. 
A potential exploitability context is shown in Appendix~\ref{apd:rsd_bugs}.}
\end{vul_ns}


\noindent\begin{vul_ns}[\label{v2}]
\noindent \textbf{(CVSS score: 8.5).} 
\rsd{} executes \textit{illegal} \texttt{LOAD} instructions, leading to DoS attacks or unauthorized access attacks.
\red{RISC-V ISA uses ``funct3'', a 3-bit field, to specify both the width of memory access and whether the load operation should be sign-extended or zero-extended. 
For example, \texttt{000} denotes a \texttt{byte} data width with sign-extended while \texttt{100} denotes a \texttt{byte} data width with zero-extended. 
No standard load instruction uses the ``funct3'' field value \texttt{111}.}
\red{A compliant processor should either raise an \textbf{illegal instruction exception} or execute them using custom data types (e.g., vector extensions). 
However, \rsd{} does not support customized data types and executes the \texttt{load} instructions with \texttt{funct3=111} without raising any exceptions. 
\rsd{} misinterprets unrecognized ``funct3'' values as byte data width \texttt{000}. 
\ourtool{} detected this vulnerability through differential testing. 
While the reference model, \textit{Spike}~\cite{spike}, correctly throws an illegal instruction exception, \rsd{} silently executes the faulty instructions. 
Listing~\ref{listing:v2_load} shows the behavior of \rsd{} and the expected behavior of \textit{Spike} in detail.}
A malicious attacker can execute instructions to compromise the system's security.
This vulnerability is an undocumented hardware feature, \texttt{CWE-1242}~\cite{hardware_cwe}.
\end{vul_ns}

\noindent\begin{vul_ns}[\label{v3}]
\noindent \textbf{(CVSS score: 8.5).} 
This vulnerability is similar to \ref{v2} except that the opcode corresponds to the \texttt{STORE} instruction. 
\red{In RISC-V, valid ``funct3'' values for \texttt{store} instructions range from \texttt{000} to \texttt{011}, corresponding to byte, halfword, and word stores.
However, \rsd{} executes the \texttt{store} instructions with other ``funct3'' field values (such as \texttt{111}) without raising any exceptions.
In contrast, \textit{spike} throws illegal instruction exceptions correctly.} 
Both vulnerabilities \ref{v2} and \ref{v3} stem from the same root cause in \rsd{}'s instruction decoding logic, as discussed in Appendix~\ref{apx:rootcause_v23}.
\blue{\ref{v2} can allow malicious programs to probe memory in ways not anticipated by the software stack, potentially enabling information disclosure. 
\ref{v3} can compromise data integrity in user-mode or shared-memory scenarios, weaken isolation boundaries, and overwrite control data structures managed by higher-privileged components. Appendix~\ref{apd:rsd_bugs} shows their potential exploitability context.}
\end{vul_ns}

\input{Table/vul_detection}

\begin{figure*}[t]
    \centering
    \includegraphics[width=\textwidth]{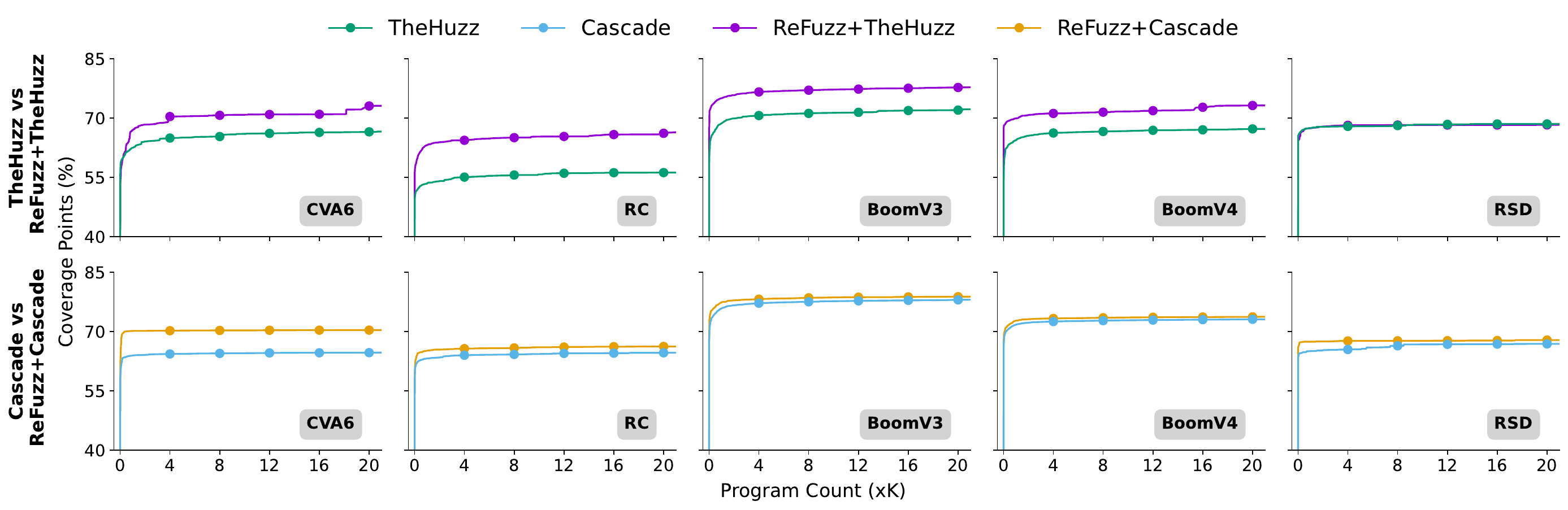}
    \caption{Branch coverage results of \thehuzz{}~\cite{kande2022thehuzz}, \cascade{}~\cite{solt2024cascade}, and \ourtool{}.}
    \label{fig:cb_vs_orig}
\end{figure*}



\blue{\subsection{Vulnerability Detection Speed}}
\blue{We compared \ourtool{}'s efficiency against \thehuzz{} and \cascade{} on vulnerability detection. 
Since \ourtool{} reuses inputs that trigger known bugs in training processors, we evaluate time-to-bug and required tests to detect new vulnerabilities in testing processors (\rsd{} and \boomf{}), which prevents overfitting.
To measure time-to-bug, we use \vcs{} as the simulation tool for all fuzzers. 
Table~\ref{tab:vul_det} summarizes the results. 
Because a wide range of instructions can trigger the two new bugs, as shown in Table~\ref{tab:tb_boom_b2} (See Appendix~\ref{apx:bugs}), their time-to-bug numbers are not informative for speed comparison and were excluded. But they still demonstrate how IP reuse can propagate vulnerabilities and how different microarchitectures can introduce vulnerability variants.}

\blue{\ourtool{} takes an average $\vulalltimespdp{}\times$ less real time and $\vulalltestspdp\times$ less tests to detect new vulnerabilities. 
\ourtool{} detects \ref{v1} faster than \thehuzz{}~\cite{kande2022thehuzz} because the seed that triggers another \texttt{FENCE.I}-related vulnerability in \cva{} is in the \textit{vulnerability list} of \ourtool{}. 
As a result, \ourtool{} does not need to generate such seeds from scratch.
\ourtool{}'s test speed is similar to \cascade{} for \ref{v1} because \cascade{} aims to generate long inputs with diverse instructions, increasing the probability of containing the instructions that can trigger \ref{v1}.
However, the time-to-bug is $\cbcascadevospdptime{}\times$ faster because the PP-test only contains the instructions that can trigger the \ref{v1}, leading to less simulation time. 
While users have to identify the instructions that trigger \ref{v1} from \cascade{}'s long inputs.
This also shows that reusing effective PP-tests can further enhance the following debugging process.}

\blue{However, \ourtool{} detected \ref{v2} and \ref{v3} that are not on the \textit{vulnerability list} slower than \thehuzz{}.
This is because \ourtool{} prioritizes the PP-tests of known vulnerability as mentioned in Section~\ref{sec:put_together}, which delays the identification of \ref{v2} and \ref{v3}. Adding the corresponding tests for \ref{v2} and \ref{v3} will improve the speed of vulnerability detection.}
\blue{Moreover, integrating \ourtool{} with \cascade{} can not detect these two vulnerabilities because \cascade{} constrains the values of the ``funct3'' field of \texttt{LOAD} and \texttt{STORE} instructions to always be in the valid range. This shows the need to include PP-tests from fuzzers with distinct mechanisms and include mutation processes to enhance design space exploration.}

\subsection{Coverage Evaluation}~\label{sec:cov_eva}
To answer \textbf{Q2}, we compare the capability of \ourtool{} in achieving coverage with the baseline fuzzers: \thehuzz{}~\cite{kande2022thehuzz} and \cascade{}~\cite{solt2024cascade}. 
Unlike the evaluations in Sections~\ref{sec:mot_cov} and \ref{sec:train_cb}, we use the PP-tests from both fuzzers to train \ourtool{} for comprehensive evaluation.
Across two testing processors, \boomf{} and \rsd{}, \ourtool{} achieves a $\cbavetestspdp{} \times$ coverage speedup on average than the baseline fuzzers. 
Specifically, \ourtool{} generates $\cbavetestspdp{} \times$ fewer tests to achieve the same \tcov{} as the baseline fuzzers.
For total coverage, \ourtool{} achieves an average of $\cbavetestincov\%$ more \tcov{}..
\red{on the testing processors.}

Though using \cva{}, \rc{}, and \boomt{} for training, \ourtool{} outperforms both baseline fuzzers in \tcov{}. 
On average, \ourtool{} achieves $\cbaveallcov{}\%$ more total coverage across all five processors, highlighting \ourtool{}'s ability to reuse highly effective tests to help fuzzers explore more design spaces. 
Figure~\ref{fig:cb_vs_orig} shows coverage results. 


\noindent\textbf{Comparison with \thehuzz{}.}
For testing processors, \ourtool{} achieves $\cbhuzzboomfinc{}\%$ more \tcov{} and a $\cbhuzzboomfspdp{} \times$ speedup compared to \thehuzz{} on \boomf{} and achieves similar total coverage ($\cbhuzzrsd{}\%$) as \thehuzz{} ($\thehuzzrsd{}\%$) on \rsd{}.
On the training processors, \ourtool{} achieves $\cbhuzzcvainc\%$ more \tcov{} on \cva{} and $\cbhuzzrcinc{}\%$ more \tcov{} on \rc{}.
This is because \ourtool{} leverages the mutation engines of \thehuzz{} to mutate tests with complex data- and control-flow logic from \cascade{}, which helps explore more design spaces than randomly generated seeds.

\noindent\textbf{Comparison with \cascade{}.}
For testing processors, \ourtool{} achieves $\cbcascadeboomfinc{}\%$ more \tcov{} and a $\cbcascadeboomfspdp{} \times$ speedup compared to \cascade{} on \boomf{}. 
On \rsd{}, \ourtool{} achieves similar \tcov{} ($\cbcascadersd{}\%$) as \cascade{} does ($\cascadersd{}\%$) but still achieves a $\cbcascadersdspdp{} \times$ \red{speedup}\blue{\spd{}} than \cascade{}.
On the training processors, \ourtool{} achieves $\cbcascadercinc{}\%$ more coverage on \rc{}.
This is because \ourtool{} incorporates tests generated by \thehuzz{}, which helps explore corner cases.
\ourtool{} achieves limited improvement on \tcov{} when integrating with \cascade{}, primarily because \cascade{} lacks a mutation engine \blue{and depends only on the existing tests}. 
\red{Without the ability to generate variants of tests, it cannot leverage highly effective tests to further explore design spaces and instead depends solely on the incremental coverage of existing tests.}

Yet, \ourtool{}  achieves $\cbcascadecvainc{}\%$ more \tcov{} and a $\cbcascadecvaspdp{}\times$ speedup over \cascade{} on \cva{}, because the \texttt{FENCE.I} will cause \cva{} to hang using \cascade{}'s setup. 
Since most of \cascade{}'s tests include this instruction, they hang early in execution, preventing subsequent instructions from running and limiting their ability to explore the design spaces of \cva{}.
\ourtool{} eliminates most of these tests during training, highlighting the importance of dropping ineffective tests and evaluating the effectiveness of tests across processors.


\noindent\textbf{Coverage improvement on \rsd{}} is limited compared to both baseline fuzzers because \rsd{} is a 32-bit RISC-V processor, whereas all training processors are 64-bit. 
As a result, many instructions of PP-tests identified during training are illegal or invalid on RSD, reducing the effectiveness of tests and hindering design space exploration. 
To address this limitation, future iterations of \ourtool{} could be trained using a set of 32-bit \riscv{} processors.


\noindent\blue{\textbf{Evaluating condition coverage}.} 
\blue{To evaluate the comprehensiveness of \ourtool{}, we also use condition coverage as the target metric.
Compared to branch coverage, condition coverage is a finer coverage metric that monitors the possible signal combinations in branch statements~\cite{chen2023hypfuzz}.}
\blue{The \ccontext{} and the adaptive threshold $\theta$ are configured as $\theta = \{45\%:14.84, 50\%:11.82, 55\%:7.81, 60\%:4.69, 65\%:2.86\}$.}
\blue{Across two testing processors, \boomf{} and \rsd{}, \ourtool{} achieves an average of \cbavetestcondspdp{}$\times$ \spd{} and \cbavetestcondincov{}$\%$ more \tcov{}. 
On average, \ourtool{} achieves a \cbaveallcondspdp{}$\times$ \spd{} and \cbaveallcondcov{}$\%$ more \tcov{} across all five processors.
Appendix~\ref{apd:cond_eva} includes more details.}


Overall, to answer \textbf{Q2}, \ourtool{} is agnostic to existing fuzzers and outperforms them in both \tcov{} and \spd{}.
These improvements highlight its ability to generate more effective tests and accelerate the exploration of processor design spaces, making it a robust and scalable solution for enhancing vulnerability detection across a wide range of processors.

\subsection{Evaluating \ourtool{}'s Optimizations and Parameters}~\label{sec:anal_dist}
To answer \textbf{Q3}, in Sections~\ref{sec:mot_cov} and \ref{sec:train_cb}, we have analyzed how \ourtool{} leverages the mutation engines of baseline fuzzers to achieve higher \tcov{}, and how the adaptive CB algorithm helps \ourtool{} identify effective tests compared to the original CB algorithm. 

In this section, we further examine how different optimizations and parameters affect \ourtool{}’s efficiency.
Specifically, we focus on the impact of the \textbf{PP-test corpus} and the number of \textbf{training steps} on the efficiency and effectiveness of the trained CB model.
Unlike other parameters such as the number of arms, which make minor adjustments to the configuration of \ourtool{}'s CB model, the PP-test corpus determines the effectiveness of tests, while the training steps decide whether the model has sufficient time to identify optimal tests.
\red{the PP-test corpus and training steps have a more general impact.}

To make a comprehensive analysis, we prepared three PP-test corpora of different sizes. The \textbf{all} corpus includes all PP-tests generated by both fuzzers. 
The \textbf{interesting} corpus includes only interesting PP-tests that achieve incremental coverage during fuzzing the prior processors.
The \textbf{minimized} corpus consists of PP-tests identified by our test minimizer.

\begin{figure}[!h]
    \centering
    \includegraphics[trim={10 4 10 0},clip,width=0.85\linewidth]{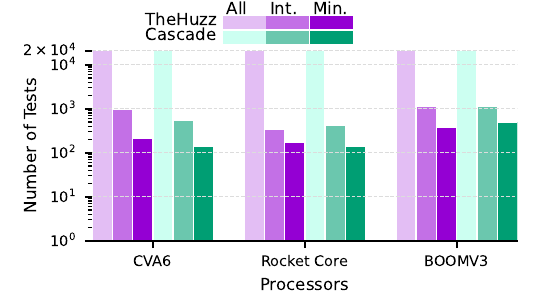}
    \caption{The number of tests the test minimizer reduces from \thehuzz{}~\cite{kande2022thehuzz} and \cascade{}~\cite{solt2024cascade}. ``Int.'' represents the interesting corpus, and ``Min.'' represents the minimized corpus.}
    \label{fig:test_reduce}
\end{figure}

\noindent\textbf{Test minimizer.}
Figure~\ref{fig:test_reduce} shows the results of applying the test minimizer to the all corpus to generate the minimized corpus.
Compared to the original 21K tests generated by each fuzzer, it significantly reduces the corpus size.
For example, it reduces the number of \thehuzz{} tests for \rc{} from 21K to 174, achieving a \thehuzzrc{} reduction rate. 
On average, the test minimizer \blue{takes \testminitime{} seconds and} achieves a \avereduction{} reduction rate compared to the number of PP-tests in the all corpus and a \aveinteretreduction{} reduction rate compared to the interesting corpus.
This substantial reduction highlights the high degree of redundancy among tests and underscores the importance of reusing highly effective tests to improve fuzzing efficiency.

\noindent\textbf{Different training steps and PP-test corpora.}
\red{As mentioned in Section~\ref{sec:why_cb}, CB algorithms are scalable to real-world problems due to their \textit{anytime learning} property, which means the longer they are trained, the more optimal solutions they will find.}
\blue{CB algorithms support \textit{anytime learning}, meaning that longer training generally yields better results.}
\red{We analyze how the number of training steps affects the effectiveness of the trained CB model. 
We evaluate six training steps, including $0.5K$, $1K$, $2K$, $5K$, $8K$, and $10K$, across three PP-test corpora.}
\blue{To study this effect, we vary the number of training steps, $0.5K, 1K, 2K, 5K, 8K$, and $10K$, and test each setting across three PP-test corpora. For every setting, we measure how many effective tests the adaptive CB algorithm identifies and repeat the experiment three times to report average results.}
\red{For each training step, we measure the number of effective tests identified by the adaptive CB algorithm and repeat each experiment three times to analyze average performance.}
\begin{figure}[!h]
    \centering
    \includegraphics[trim={2 0 2 1},clip,width=0.95\linewidth]{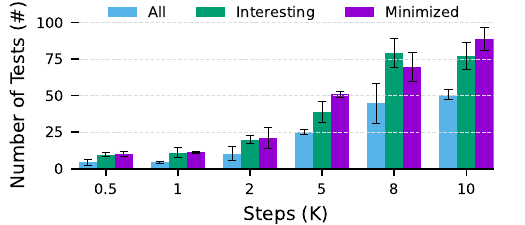}
    \caption{Training \ourtool{}'s CB model with different training steps.}
    \label{fig:train_dif_steps}
\end{figure}

Figure~\ref{fig:train_dif_steps} presents the average number of effective tests on the coverage list when the coverage context is $70\%$. 
Using the all corpus as an example, we observe that as training steps increase from $0.5K$ to $10K$, the average number of effective tests rises from $4$ to $51$, demonstrating the \textit{anytime} property of CB algorithms.
Because the minimized corpus contains the smallest subset of PP-tests that achieves equivalent total coverage as the all corpus, \ourtool{} is able to identify more effective tests from it under the same training steps. 
For example, at $10K$ training steps, \ourtool{} identifies an average of $89$ effective tests from the minimized corpus, compared to $77$ from the interesting corpus and $51$ from the all corpus. 
This result underscores the impact of the test minimizer in improving the training efficiency and the CB model's effectiveness.

Compared to other coverage contexts, identifying the effective tests for the $70\%$ coverage context requires longer training steps, as the remaining uncovered points tend to be corner cases. 
\ourtool{} requires fewer training steps to find effective tests for other coverage contexts. 
\red{Further details across all coverage contexts are provided in Appendix~\ref{apd:train_step}.}

To answer \textbf{Q3}, the test minimizer effectively reduces the size of the PP-test corpus while preserving total coverage, demonstrating its value in eliminating redundant tests. 
Analysis of different training steps confirms the anytime property of CB algorithms and indicates that highly effective tests identified by the test minimizer can enhance the efficiency of the training process.

%% file: Table/vul_detection.tex
\begin{table*}[h]
\centering
\caption{Summary of Vulnerability Detection Speed of \thehuzz{}~\cite{kande2022thehuzz}, \cascade{}~\cite{solt2024cascade}, and \ourtool{}. N.D. refers ``Not Detected."}
\label{tab:vul_det}
\resizebox{0.95\textwidth}{!}{%
\begin{tabular}{ccccccccccccccc}
\hline
\multirow{2}{*}{\textbf{Vulnerability}} &
  \multirow{2}{*}{\textbf{Processor}} &
  \multirow{2}{*}{\textbf{CWE}} &
  \multicolumn{6}{c}{\textbf{Time (sec)}} &
  \multicolumn{6}{c}{\textbf{Tests}} \\ \cline{4-15} 
 &
   &
   &
  \thehuzz{} &
  \begin{tabular}[c]{@{}c@{}}\ourtool{}+\\ \thehuzz{}\end{tabular} &
  Speedup &
  \cascade{} &
  \begin{tabular}[c]{@{}c@{}}\ourtool{}+\\ \cascade{}\end{tabular} &
  Speedup &
  \thehuzz{} &
  \begin{tabular}[c]{@{}c@{}}\ourtool{}+\\ \thehuzz{}\end{tabular} &
  Speedup &
  \cascade{} &
  \begin{tabular}[c]{@{}c@{}}\ourtool{}+\\ \cascade{}\end{tabular} &
  Speedup \\ \hline
\begin{tabular}[c]{@{}c@{}}V1: \texttt{FENCE.I} causes\\ memory deadlock\end{tabular} &
  \rsd{} &
833   &
\thehuzzvoavetime{}  &
\cbthehuzzvoavetime{}   &
$\cbthehuzzvospdptime{}\times$   &
\cascadevoavetime{}   &
\cbcascadevoavetime{}   &
$\cbcascadevospdptime{}\times$   &
\thehuzzvoavetest{}   &
\cbthehuzzvoavetest{}   &
$\cbthehuzzvospdptest{}\times$  &
\cascadevoavetest{}   &
\cbcascadevoavetest{}   &
$\cbcascadevospduptest{}\times$
   \\ \hline
\begin{tabular}[c]{@{}c@{}}V2: Illegal \texttt{LOAD}\\ can be executed\end{tabular} &
  \rsd{} &
  1242 &
\thehuzzvtwavetime{}  &
\cbthehuzzvtwavetime{}   &
$\cbthehuzzvtwspdptime{}\times$   &
N.D. &
N.D. &
---&
\thehuzzvtwavetest{}   &
\cbthehuzzvtwavetest{}   &
$\cbthehuzzvtwspdptest{}\times$  &
N.D.   &
N.D.   &
---
   \\ \hline
\begin{tabular}[c]{@{}c@{}}V3: Illegal \texttt{STORE}\\ can be executed\end{tabular} &
  \rsd{} &
  1242 &
\thehuzzvtavetime{}  &
\cbthehuzzvtavetime{}   &
$\cbthehuzzvtspdptime{}\times$   &
N.D.   &
N.D.  &
---  &
\thehuzzvtavetest{}   &
\cbthehuzzvtavetest{}   &
$\cbthehuzzvtspdptest{}\times$  &
N.D.   &
N.D.  &
---
   \\ \hline
\end{tabular}
}
\end{table*}

%% file: draft/relatedwork.tex
\section{Related Work}\label{sec:re_work}
\red{In recent years, processor fuzzing has become an active area of research, with efforts focusing primarily on three core challenges: generating high-quality tests, designing effective coverage metrics to guide test generation, and improving vulnerability detection capabilities.}
\red{Early hardware fuzzers~\cite{hur2021difuzzrtl,canakci2022processorfuzz,kande2022thehuzz} relied on random input generation, generic coverage metrics, and traditional differential testing for vulnerability detection. 
More recent work has introduced domain-specific enhancements, including constrained input generation~\cite{solt2024cascade,xu2023morfuzz}, formal-assisted seed generation~\cite{chen2023hypfuzz}, machine learning-based test generation~\cite{rostami2024chatfuzz,wugenhuzz,wu2025hfl,gotz2025rlfuzz}, and custom feedback mechanisms for vulnerability detection~\cite{specure,borkar2024whisperfuzz}.}

\noindent \textbf{Seed generation.} Traditional approaches, such as those used in \difuzz{}\cite{hur2021difuzzrtl} and \profuzz{}\cite{canakci2022processorfuzz}, rely on randomly generating instruction sequences~\cite{hur2021difuzzrtl,kande2022thehuzz,sugiyama2023surgefuzz,fuzzhwlikesw}. 
\red{While simple, this often results in low-quality, semantically invalid inputs that limit coverage and bug-finding performance. 
To improve seed complexity and relevance, recent tools have proposed constrained instruction generation.} 
\morfuzz{}\cite{xu2023morfuzz} and \cascade{}\cite{solt2024cascade} increase input diversity by enforcing control- and data-flow constraints during generation. 
\red{However, these methods rely heavily on manual design effort and may over-constrain the instruction space.} 
\red{Formal methods such as }\hypfuzz{}~\cite{chen2023hypfuzz}\red{ take a different approach by using symbolic execution and} \blue{uses} formal verification to synthesize inputs that reach hard-to-reach design spaces\blue{ but scale poorly with complex designs}. 
\red{While powerful, these techniques are computationally expensive and scale poorly with complex designs.} 
Machine learning-based approaches~\cite{rostami2024chatfuzz,gotz2025rlfuzz}, such as \texttt{HFL}\cite{wu2025hfl} and \genhuzz{}~\cite{wugenhuzz}, explore the use of large language models or reinforcement learning to generate instruction streams with semantic and structural dependencies. 
Despite promising early results, these methods often suffer from hallucinations or invalid tests, limiting their practicality.

\noindent\blue{\textbf{Seed schedule.} 
\mabfuzz{}~\cite{gohil2024mabfuzz} leverages multi-armed bandit approaches and coverage feedback to identify an optimal schedule of seeds for each PUT.
However, it is incompatible with non-feedback-based fuzzers like \cascade{} and relies on the baseline fuzzer's seed-generation strategy.
\ourtool{} discards fuzzers' internal strategies and intelligently uses effective PP-tests to enhance fuzzing efficiency.}

\noindent \textbf{Coverage metrics.} Existing processor fuzzers either rely on industry-standard code coverage metrics~\cite{kande2022thehuzz,chen2023hypfuzz,wugenhuzz,wu2025hfl,rostami2024chatfuzz} or define custom hardware-specific metrics~\cite{rfuzz,hur2021difuzzrtl,xu2023morfuzz,canakci2022processorfuzz}. Standard coverages are broadly compatible with existing design flows and simulation environments widely used in industry~\cite{vcs_companies_enlyft,vcs_companies_theirstack}, making them the most practical option for testing production-grade designs. 
Custom coverage metrics attempt to capture more internal behaviors, such as multiplexer activity~\cite{rfuzz}, register toggling~\cite{hur2021difuzzrtl}, or instruction semantics~\cite{xu2023morfuzz} \blue{ but face integration barriers to industry verification flows.}
\red{While informative, these methods often rely on access to internal signals or require designs to be written in specific hardware description languages (HDLs) like \textit{Chisel}\cite{bachrach2012chisel}, limiting their applicability. 
As most real-world designs are implemented in \textit{Verilog} or \textit{SystemVerilog}\cite{piziali2008functional,verilog_report_siemens}, such approaches face significant integration barriers.}
\blue{\ourtool{} adopts industry-standard code coverage and ensures simple integration with industry verification flows.}
\red{\ourtool{} adopts industry-standard code coverage as its feedback mechanism, ensuring compatibility with widely used simulation tools. }

\noindent \textbf{Vulnerability detection.} 
Most processor fuzzers are designed to detect functional bugs through differential tests~\cite{hur2021difuzzrtl,canakci2022processorfuzz,kande2022thehuzz,rostami2024chatfuzz, wugenhuzz, chen2023hypfuzz, gotz2025rlfuzz}. 
Orthogonal efforts target microarchitectural vulnerabilities, including speculative execution attacks~\cite{hur2022specdoctor,ghaniyoun2021introspectre,specure,de2025phantom} and timing side-channels~\cite{borkar2024whisperfuzz,rajapaksha2023sigfuzz}\blue{, leveraging techniques like information flow tracking (IFT)~\cite{hu2021hardware,solt2022cellift}}. These works differ in their threat models and detection methods but share the same underlying need for effective test generation.
Since \ourtool{} focuses on optimizing\red{ test generation} \blue{test reuse}, it is agnostic to the target vulnerability and can be integrated with a wide range of detection strategies.


\blue{Overall, \ourtool{} fills a key gap in prior work, which treats each processor independently and generates seeds from scratch, ignoring useful information from earlier testing efforts. 
\ourtool{} introduces a CB-based test reuse framework that adaptively selects and mutates tests from prior processors for the current PUT. 
This approach improves coverage and increases the likelihood of uncovering cross-generational vulnerabilities and their variants.}
\red{Overall, \ourtool{} addresses a gap left by prior work: treat each processor in isolation, generating seeds from scratch and discarding valuable insights from previous testing campaigns. 
Our work addresses this limitation by introducing a novel test reuse framework based on contextual bandit algorithms. The algorithms adaptively select tests from prior processors, allowing them to be effectively reused and mutated for processor-under-tests. \ourtool{} not only improves coverage but also increases the chance of detecting cross-generational vulnerabilities and their variants.}

%% file: draft/conclusion.tex
\section{Discussion}~\label{sec:discuss}

\noindent\blue{\textbf{Integrating \ourtool{} with industry verification pipelines.}
\ourtool{} naturally fits into the continuous integration and regression testing pipeline for IP vendors due to its \textit{anytime learning} property, allowing \ourtool{} to continuously identify and reuse effective PP-tests while keeping the trained model within a reasonable size.
Moreover, besides \textit{Synopsys} \vcs{}~\cite{vcs}, \ourtool{} is compatible with coverage metrics from other industrial simulators, such as \textit{Cadence} \xce{}~\cite{xce} and \textit{Siemens} \texttt{Questa One Sim}~\cite{questa}, which also generate coverage files similar to \vcs{}'s coverage database.}

\noindent\blue{\textbf{Future challenges.} 
To further strengthen \ourtool{}’s impact on the semiconductor development cycle, we identify three key challenges:}
\blue{\textbf{(i)~Generalizability.}
In practice, hardware designs vary in ISA extensions (e.g., 32-bit vs. 64-bit \riscv{}, or cross-ISA settings such as \riscv{} and \arm{}) and verification platforms (e.g., FPGA-prototyping, black-box, and hybrid flows).}
\blue{\textbf{(ii) Sparsity.} 
Reuse may be constrained when vendors have limited historical designs or employ heavily customized architectures.}
\blue{And \textbf{(iii) Scalability.} 
The reliance on integer programming to build the test minimizer may raise scalability concerns when applied to large industrial test suites.}

\noindent\blue{\textbf{Future directions.}
For generalizability, \ourtool{} can evolve in two directions.
First, it can support more contexts, such as ISA extensions or even cross-ISA scenarios, so that it can reuse more precise tests, similar to selecting ISA extensions when compiling binaries.
Second, an abstract CB model could use microarchitectural features, functionalities, and ISA extensions as context, along with mappings between instructions and the features they exercise. Under a given context, the model could identify the relevant features and their interactions and then generate appropriate inputs.}
\blue{
Finally, \ourtool{} can adapt to verification platforms with different observability levels by using any quantitative metric that defines 0\% and 100\% coverage. As long as coverage increments can be computed (Definition~\ref{defn_3}), \ourtool{} can operate on that platform.}

\blue{For sparsity, when historical design data are limited, vendors can enrich their test suites by incorporating tests from other ISAs and translate them to the target ISA. 
For designs with highly customized or sparse architectures, integrating additional mutation engines into \ourtool{} can help explore architecture-specific features.}

\blue{Finally, for scalability, combining integer linear programming (ILP), clustering, and approximation algorithms can tackle the challenge. 
One strategy is to partition the overall suite into smaller subsets and build a hierarchical approach.
The approaches perform ILP on each subset, merge the resulting minimal suites, and then run ILP again to obtain a global minimal set. This approach, coupled with parallelization, can substantially improve efficiency. 
If scalability issues persist even after clustering, approximation algorithms, such as greedy heuristics, can provide near-optimal solutions with significantly lower overhead.} 

\section{Conclusion}\label{sec:conclu}
\red{Modern processor designs rely on iterative development and design reuse, leading to hardware vulnerabilities that propagate across processor generations.}
\blue{Design reuse leads to vulnerabilities that propagate across processor generations.}
\blue{\ourtool{} proposes the first test reuse hardware fuzzing framework that reuse and mutate effective tests from prior processors to enhance fuzzing on processors-under-test, leveraging contextual bandit algorithms.}
\ourtool{} is also agnostic to processor fuzzers or any dynamic techniques that require seeds to initiate their processes.
\red{It improves both vulnerability detection and coverage, detecting new bugs across multiple RISC-V processors and highlighting the risks of design reuse.} 
\ourtool{} uncovered three new vulnerabilities.
One was triggered by the same test that triggered vulnerabilities in a prior processor. 
\blue{It detects two functional bugs across three processors due to shared designs.}
Evaluations show that \ourtool{} achieves an average $\cbavetestspdp{}\times$ \spd{} and $\cbavetestincov{}\%$ more \tcov{} over baseline fuzzers. 
These results establish \ourtool{} as a practical solution for\red{ pre-silicon} processor verification, especially for companies like \textit{Intel} and \textit{AMD}, which have effective tests from a broad range of prior processors.
\red{, demonstrating that processors can contain different vulnerabilities when implementing the same functionalities.} 
\red{Additionally, \ourtool{} detects two functional bugs across three processors, showing how design reuse can propagate vulnerabilities across processors, and how new microarchitectures can introduce variants of known vulnerabilities.}

\red{\noindent\textbf{How to use other fuzzers?}
As mentioned in Section~\ref{sec:implementation}, we trained the CB model in \ourtool{} using \thehuzz{} and \cascade{} as baseline fuzzers. 
However, \ourtool{} is not limited to these fuzzers. 
To integrate a new fuzzer into \ourtool{}, one simply needs to follow the training procedure described in Section~\ref{sec:train}: run the new fuzzer on a set of target cores, collect the corresponding coverage tests, and record the tests that trigger potential vulnerabilities. These tests are then passed through our test minimizer to reduce redundancy. Finally, coverage contexts are sampled and used to retrain the CB model. Our framework automates all steps beyond the initial setup of the new fuzzer, making it straightforward to incorporate and benefit from future fuzzing techniques.}

\noindent\red{\textbf{How to use other coverage metrics?}
In this work, we used branch coverage as the target metric due to its demonstrated effectiveness in uncovering hardware vulnerabilities~\cite{chen2023hypfuzz}, as described in Section~\ref{sec:train}. 
However, the \ourtool{} framework is not limited to branch coverage. 
Users can configure the CB algorithm to optimize for any coverage metrics supported by commercial simulation tools, depending on their verification goals. 
Moreover, \ourtool{} supports the use of multiple coverage metrics simultaneously, allowing users to assign custom weights to combine them as needed. 
This flexibility is fully supported by the design of the \ourtool{} framework.}

\red{\noindent\textbf{Contexts with more relevant information.}
As shown in the fuzzing results for \rsd{} in Section~\ref{sec:cov_eva}, processors following the same ISA can have different ISA extensions. 
Including such information in the context can enhance test selection. For example, if a processor lacks support for vector operations, explicitly encoding the absence of relevant ISA extensions helps the CB model avoid selecting tests that include those instructions---analogous to specifying ISA extensions during program compilation to ensure compatibility with the PUT.}

\section{Acknowledgement}
We thank Stephen Muttathil for his contribution to \cascade{} implementation.
Our research work was partially funded by Intel's Scalable Assurance Program, 
the US Office of Naval Research (ONR Award \#N00014-18-1-2058),
the Lockheed Martin Corporation,
the SCALE Fellowship Program,
DFG-SFB 1119-236615297, the European Union under Horizon Europe Programme-Grant Agreement 101070537-CrossCon, NSF-DFG-Grant 538883423, and the European Research Council under the ERC Programme-Grant 101055025-HYDRANOS. 


%% file: draft/ethic.tex
\section{Ethics Considerations}
We consider the paper without potentially negative outcomes from tangible harms and violations of human rights.
We consider the full spectrum of stakeholders and the paper, including no harms related to exposing the identity of research subjects, such as the individual identities, groups, and organizations, and the behaviors, communications, or relationships associated with such identification. 
\noindent\textbf{Respect for Persons.} The research targets open-sourced hardware designs and does not involve natural persons, live systems, or certain data that identifies them. Thus, no harm is caused to human subjects, non-subjects, and information and communication technology users, such as disruption of access, loss of privacy, or unreasonable constraints on protected speech or activities. 
\noindent\textbf{Beneficence.} To minimize the relevant harmful impacts, we contacted the vendors of open-sourced benchmarks as soon as our strategy detected the vulnerabilities. 
Moreover, based on our current knowledge, the vulnerabilities exist in an open-source benchmark, which has not been used for any commercial purposes. 
We discuss the root causes of vulnerabilities in the paper to help vendors and potential users mitigate the vulnerabilities. 
\noindent\textbf{Justice: Fairness and Equity.} 
We select benchmarks that are widely used in existing papers. We select existing techniques as baselines based on their distinct and unique features. We evaluate the performance of our technique and baselines on the same platform for a fair comparison. Based on our best understanding of existing techniques, we skip comparisons that may lead to unfair comparisons.

%% file: draft/appendix.tex
\section*{Appendix}
\setcounter{section}{0}
\renewcommand{\thesection}{\Alph{section}}

\titleformat{\section}
  {\normalfont\raggedright} 
  {\thesection}{1em}{}   

\red{Algorithm~\ref{algo_cb} outlines the $k$-armed contextual bandit.
In each round $t$, the learner observes a context $c_t$, selects an action $a_t$ based on the policy $\pi$, and receives a reward $r(c_t, a_t)$. 
For example, in recommending movies, $r(c_t,a_t)$ represents the score on the movie $a_t$ given by the user $c_t$.} 

\section{Optimization Model} \label{apx:appendix_model}
This section shows the optimization model for the test minimizer as mentioned in Section~\ref{sec:train_cb}.

Consider the following notations:
\begin{enumerate}
    \item $\mathbf{T}$: coverage matrix $\mathbf{T}\coloneqq[t_{ij}]$, where $t_{ij}$ is a binary indicator that represents whether the coverage point $j$ has been covered ($1$) or not ($0$) in test $i$. 
    \item $\mathbf{L_j}$: minimum coverage constraint for coverage point $j$. 
    \item $\mathbf{U_j}$: maximum coverage constraint for coverage point $j$. 
    \item $\mathbf{x_i}$: $x_i \in {0,1}, \forall i$, a decision variable indicating if test $i$ is in the minimal subset. 
\end{enumerate}

\noindent The model is then formulated as:
\begin{enumerate}
    \item \textbf{Objective Function}: Identify the minimal subset of tests that maintains the coverage equivalent: $\textbf{minimize}\quad \sum_{i}x_i$.
    \item \textbf{Constraint 1}: Ignore coverage point $j$ if it is always covered in all tests: $\sum_{i}t_{ij}x_i = 0$.
    \item \textbf{Constraint 2}: The constraint for the rest of the coverage points to identify the minimal subset of tests: $L_j \leq \sum_{i}t_{ij}x_i \leq U_j,$
    where If at least one test cover point $j$, we let $L_j = 1$ and $U_j = \infty .$ If no tests cover point $j$, we let $L_j = U_j = 0.$
\end{enumerate}

\makeatletter
\let\@ORGmakecaption\@makecaption
\long\def\@makecaption#1#2{\@ORGmakecaption{#1}{#2}\vskip\belowcaptionskip\relax}
\makeatother

\section{\blue{Fine-Tune Adaptive Thresholds Automatically}} \label{apx:fine_tune_threshold}

\blue{The configuration of the adaptive threshold $\theta$ is automated, as shown in Algorithm~\ref{alg:fine_tune}.
The goal is to identify approximately the same number of tests, defined by the number of arms $k$, for each \clist.
The inputs are the PP-test corpus $\mathcal{A}_{\text{corpus}}$, \ccontext{}s $\mathcal{C}$, the number of arms $k$, the check window $\gamma$, the training step $n$, and the tolerance factor $f$. 
The output is the fine-tuned adaptive thresholds $\theta$.
}

\blue{
The algorithm first calculates the acceptable interval for the number of tests (Line 3). 
Then, for each \ccontext{}, it starts the binary search from $0.00$, assuming all tests are effective, and $100.00$ means a test can cover all points of a training processor (Lines 6 to 11).
Given the threshold range with a precision of $0.01$, the binary search requires at most $14$ iterations.
To avoid duplicate tests across coverage lists, the algorithm runs our adaptive CB algorithm once, removes the chosen tests from the PP-test corpus (Lines 13 and 14), and then fine-tunes the threshold for the next \ccontext{}.
}

\blue{Note that some tests may not have coverage results during fine-tuning.
To avoid repeated simulations, we simulate the test once and reuse its coverage result.
For experimental purposes, we simulate all \mintot{} tests in parallel after test minimization using 10 threads.
The same number of threads is used during evaluation.
The total simulation time is approximately \totmintesthr{} hours in real time. 
Algorithm~\ref{alg:fine_tune} then took around \finetunetimehr{} hours to identify the thresholds.
This fine-tuning stage illustrates the necessity of test minimization and can be further accelerated with additional threads or more advanced parallelization.}

\blue{To automatically fine-tune more parameters, such as \ccontext{}s and check window $\gamma$, future work can consider involving more advanced algorithms, such as hierarchical contextual bandits~\cite{hong2022hierarchical}.}
\input{Code/config_theta}

\section{Details on Detected Bugs} \label{apx:bugs}

\noindent\begin{bug_ns}[\label{b1}]\textbf{.}
in \rc{}~\cite{rocket_chip_generator}, \boomt{}, and \boomf{}~\cite{boom}, the
\texttt{ECALL} and \texttt{EBREAK} instructions incorrectly increase the value of the \texttt{minstret} register, violating the RISC-V specification.
According to the RISC-V ISA, these exception-generating instructions should not contribute to the instruction retirement count, and\red{ thus} should leave \texttt{minstret} unchanged. 
This bug misinterprets performance counter values and constitutes a violation of expected behavior (CWE-440)~\cite{hardware_cwe}.

The bug exists across three processors due to \textbf{design reuse} and characteristics of the \textit{Chisel} hardware programming language~\cite{bachrach2012chisel}. 
Specifically, \rc{} has a module called \texttt{CSR}, responsible for updating the values of control and status registers (CSRs), including \texttt{minstret}. 
\boomt{} and \boomf{} use the same module, thereby propagating the same bug. This bug also shows that the widely applied IP reuse strategy in hardware design can potentially propagate vulnerabilities to multiple processors.
\end{bug_ns}

\noindent\begin{bug_ns}[\label{b2}]\textbf{.}
In \boomt{}, and \boomf{}~\cite{boom}, multiple instructions fail to correctly increase the value of \texttt{minstret} register.
For example, when executing the \texttt{MUL} instruction in \boomf{}, \texttt{minstret} register increments twice instead of once as expected. 
However, the same bug does not exist in \boomt{} and \rc{}, as discussed in Section~\ref{sec:motivation}.
Though three processors use the same \texttt{CSR} module, they can include bug variants due to different microarchitectures. Table~\ref{tab:tb_boom_b2} shows the common instructions that can trigger this bug in \boomt{} and \boomf{} and the additional instructions that can trigger this bug only in \boomf{}.
\input{Table/boom_b2_v3}
\end{bug_ns}

\section{\blue{Potential Exploitability Contexts of \rsd{} Vulnerabilities}}~\label{apd:rsd_bugs}
\blue{\ref{v1} presents a denial-of-service vector. Any user-mode program capable of executing arbitrary instructions can intentionally invoke \texttt{FENCE.I} to cause the memory deadlock of the processor. This makes the vulnerability practically exploitable on systems where untrusted or partially trusted code is allowed, which interrupts critical tasks, disrupts real-time workloads, and forces a hardware reset. A potential exploitability context is shown in Appendix~\ref{apd:rsd_bugs}.}
\blue{Attackers can exploit \ref{v1} as shown in Listing~\ref{listing:v1_fencei_mr}. This exploit demonstrates that issuing a program with \texttt{FENCE.I} instruction on the vulnerable \rsd{} processor can trigger internal deadlock. 
The program executes \texttt{FENCE.I} directly in user mode without requiring any privileged operations, allowing an attacker to reliably induce a denial-of-service attack on the system.}

\begin{lstlisting}[language=Verilog, label = {listing:v1_fencei_mr}, caption={\texttt{FENCE.I} causes memory deadlock on \rsd{}~\cite{mashimo2019open}.},style=prettyverilog,belowcaptionskip=4pt,belowskip=5pt,aboveskip=0pt,firstnumber=1,linewidth=0.95\linewidth,xleftmargin=12pt]
int main(void) {
    // Execute FENCE.I instruction
    asm volatile ("fence.i" ::: "memory");
    return 0;
}
// RSD outputs on the debug console:
//   Deadlock detected
//   ...
//   Deadlock detected
//   MSHR deadlock detected
\end{lstlisting}


\blue{\ref{v2} can allow malicious programs to probe memory in ways not anticipated by the software stack, potentially enabling information disclosure. While this behavior alone does not inherently grant privilege escalation, it can weaken memory-safety assumptions in system software, including bounds-checking frameworks that rely on predictable exception semantics.
\ref{v3} can compromise data integrity in user-mode or shared-memory scenarios, weaken isolation boundaries, and potentially overwrite control data structures managed by higher-privileged components.}

\blue{Attackers can exploit \ref{v2} and \ref{v3} as shown in Listing~\ref{listing:v23_loadstore}. 
The attacker uses the same illegal load/store encoding (funct3 = 111) to perform unintended sign-extended byte operations. 
While the expected behavior is for the processor to throw exceptions, \rsd{} internally sign-extends the accessed byte and completes the operation as though it were a legal instruction. 
Because the extension semantics differ from the ISA specification, the resulting value in the destination register or memory location may deviate significantly from the expected result. 
This behavior enables data corruption during program execution and breaks data integrity.}

\begin{lstlisting}[language=Verilog,label={listing:v23_loadstore}, caption={\rsd{}~\cite{mashimo2019open} executes \textit{illegal} \texttt{LOAD} or \texttt{STORE} instructions.},style=prettyverilog,belowcaptionskip=4pt,belowskip=3pt,aboveskip=0pt,firstnumber=1,linewidth=0.95\linewidth,xleftmargin=12pt]
volatile uint32_t victim_mem = 0x12345678;
int main(void) {
    uint32_t tmp = 0;
    // Execute STORE instruction with illegal funct3
    asm volatile (
        ".word 0x01DE7023 // illegal STORE (funct3 = 111), rs2 = x29 (t4), rs1 = x28 (t3)"
        :
        : "r"(tmp), "m"(victim_mem)
    );
    // Execute LOAD instruction with illegal funct3
    asm volatile (
        ".word 0x000E7E83 // illegal LOAD (funct3 = 111), rd = x29 (t4), rs1 = x28 (t3)"
        : "=r"(tmp)
        : "0"(tmp), "m"(victim_mem)
    );
    return 0;}
\end{lstlisting}



\section{Root Cause of Illegal \red{Load}\blue{\texttt{LOAD}}\&\red{Store}\blue{\texttt{STORE}} Instructions in \rsd{}}\label{apx:rootcause_v23}
Listing~\ref{listing:rootcause} shows the root cause of vulnerabilities~\ref{v2} and \ref{v3} in \rsd{}'s decoding logic\cite{mashimo2019open}. 
The \texttt{case} statement decides the data width based on the value of ``funct3''. However, since \rsd{} has a \texttt{default} statement, it identifies any unrecognized funct3 values as the byte data width with sign-extended (Lines 725--727). The mitigation can add illegal instruction exception for invalid values. 
\bgroup
        \createlinenumber{1}{704}
        \createlinenumber{2}{705}
        \createlinenumber{3}{706}
        \createlinenumber{4}{707}
        \createlinenumber{5}{708}
        \createlinenumber{6}{709}
        \createlinenumber{7}{710}
        \createlinenumber{8}{711}
        \createlinenumber{9}{712}
        \createlinenumber{10}{\dots}
        \createlinenumber{11}{725}
        \createlinenumber{12}{726}
        \createlinenumber{13}{727}
        \createlinenumber{14}{\dots}
\begin{lstlisting}[language=Verilog, label = {listing:rootcause}, caption={The root cause of illegal load and store instructions.},style=prettyverilog,belowcaptionskip=4pt,belowskip=5pt,aboveskip=0pt,firstnumber=1,linewidth=0.98\linewidth,xleftmargin=12pt]
case(funct3)
    MEM_FUNCT3_SIGNED_BYTE : begin  //LB, SB
        memAccessMode.isSigned = TRUE;
        memAccessMode.size = MEM_ACCESS_SIZE_BYTE;
    end
    MEM_FUNCT3_SIGNED_HALF_WORD : begin  //LH, SH
        memAccessMode.isSigned = TRUE;
        memAccessMode.size = MEM_ACCESS_SIZE_HALF_WORD;
    end
...
    default : begin
        memAccessMode.isSigned = TRUE;
        memAccessMode.size = MEM_ACCESS_SIZE_BYTE;
...
\end{lstlisting}
\egroup

\section{Fuzzer Selection}~\label{apd:fuzz}
Existing hardware fuzzers fall into two categories: feedback-based and non-feedback-based fuzzers~\cite{rostami2024fuzzerfly}.
Feedback-based fuzzers typically use code coverage~\cite{kande2022thehuzz,chen2023hypfuzz,wugenhuzz,wu2025hfl} or customized metrics~\cite{rfuzz,hur2021difuzzrtl,xu2023morfuzz,canakci2022processorfuzz} to guide exploration of design spaces.
In contrast, non-feedback-based fuzzers~\cite{solt2024cascade} constrain input spaces based on ISA and processor design routines, introducing random behaviors, such as the random selection of operands and opcodes, to explore a broad range of design spaces. 

Feedback-based fuzzers have demonstrated effectiveness in detecting vulnerabilities, but they face compatibility issues with processor designs.
For example, \rfuzz{}~\cite{rfuzz} and \difuzz{}~\cite{hur2021difuzzrtl} customize coverage metrics for the toggling of multiplexer signals and control registers, respectively.
However, the instrumentation requires designs to be written in \textit{Chisel}~\cite{bachrach2012chisel}, while over $80\%$ of hardware designs in industry use \textit{Verilog} or \textit{SystemVerilog}, which are not supported by these fuzzers~\cite{piziali2008functional,verilog_report_siemens}.
\profuzz{} monitors control and status registers (CSRs) as coverage feedback, but such registers are outputs and do not directly reflect the explored design spaces.
As a result, deploying feedback-based fuzzers with customized coverage metrics is incompatible with most processor designs.
Nonetheless, since software simulators with code coverage metrics are widely used in semiconductor companies~\cite{vcs_companies_enlyft, vcs_companies_theirstack}, fuzzers using code coverage metrics are compatible with a broader range of processor designs that share the same ISA~\cite{kande2022thehuzz,wugenhuzz,xu2023morfuzz}.

Finally, we select the feedback-based fuzzer, \thehuzz{}~\cite{kande2022thehuzz} and the non-feedback-based fuzzer \cascade{}~\cite{solt2024cascade} as the baseline fuzzers due to their distinct mechanisms and generalizability. 
\thehuzz{} uses industrial standard code coverage metrics~\cite{vcs} and deploys AFL-like mutators, similar to those used in software fuzzers~\cite{citeafl}, which mutate seeds with limited or no knowledge of ISAs.
It generates tests that contain 20 instructions, which may limit its ability to detect vulnerabilities that require longer instructions, but enables it to explore corner cases due to its randomness.
In contrast, \cascade{}~\cite{solt2024cascade} constrains the instruction space based on ISA and generates tests of long instruction sequences with complex data- and control-flows. 
However, \cascade{} may over-constrain the randomness of instruction generation, \blue{missing corner cases.}\red{causing it to miss some corner cases.}

\red{This section shows the details of how different training steps and PP-test corpora can impact the effectiveness of the training stage across all coverage contexts. Figures~\ref{fig:train_dif_steps_all},~\ref{fig:train_dif_steps_interesting}, and~\ref{fig:train_dif_steps_minimized} show the number of effective tests identified during training using the \textbf{all}, \textbf{interesting}, and \textbf{minimized} corpora, respectively. }

\red{As discussed in Section~\ref{sec:anal_dist}, identifying effective tests for the $70\%$ coverage context requires more training steps than other contexts, since the remaining uncovered points are typically corner cases. 
In contrast, for coverage contexts of $55\%$, $60\%$, and $65\%$, \ourtool{} can identify effective tests within just $2K$ training steps, demonstrating the efficiency of our contextual bandit (CB) model.
For coverage lists of $55\%$, $60\%$, and $65\%$ coverage contexts, \ourtool{} can identify effective tests within $2K$ training steps, highlighting the efficiency of our CB model.}






\begin{figure*}[!htb]
    \centering
    \includegraphics[width=0.9\textwidth]{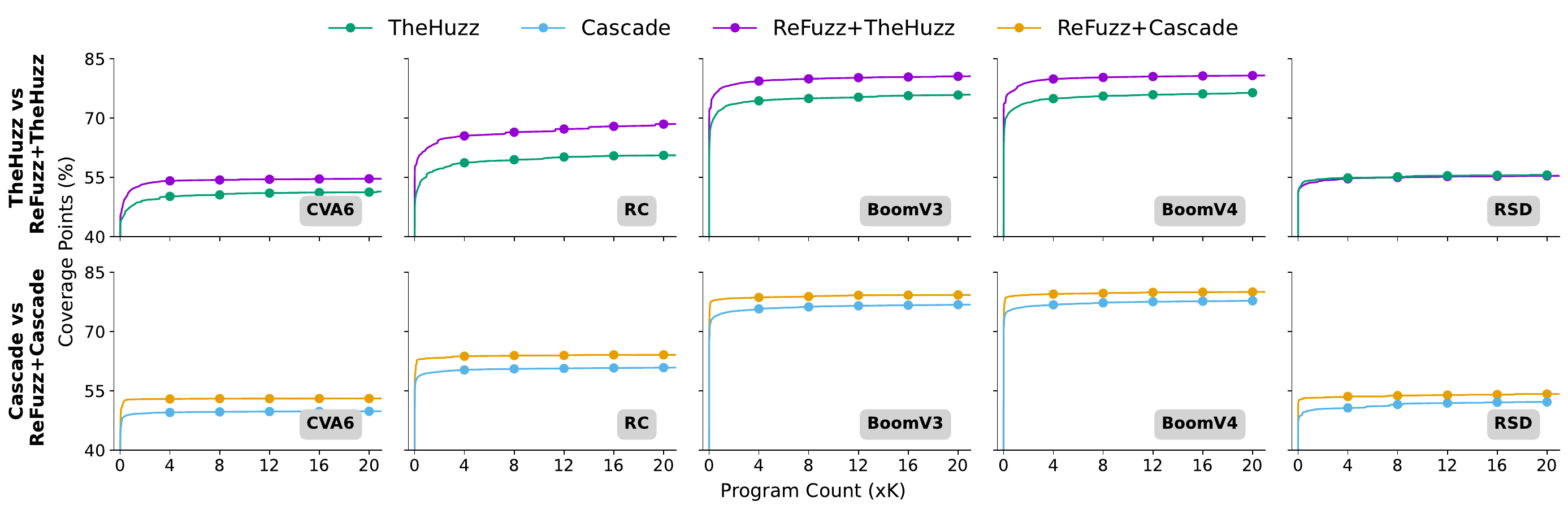}
    \caption{Condition coverage results of \thehuzz{}~\cite{kande2022thehuzz}, \cascade{}~\cite{solt2024cascade}, and \ourtool{}.}
    \label{fig:cb_vs_orig_cond}
\end{figure*}

\section{\blue{Evaluation on Condition Coverage Metric}}\label{apd:cond_eva}
\blue{We evaluated \ourtool{}'s generalized improvement using \textit{condition} coverage, including both \spd{} and \tcov{} improvement based on the number of tests (Program Count). 
Figure~\ref{fig:cb_vs_orig_cond} shows condition coverage results similar to Figure~\ref{fig:cb_vs_orig} for branch coverage.
The \ccontext{} and the adaptive threshold are configured as $\theta = \{45\%:14.84, 50\%:11.82, 55\%:7.81, 60\%:4.69, 65\%:2.86\}$.
\ourtool{} outperforms both \thehuzz{} and \cascade{} on the condition coverage, showing its comprehensiveness.
Across two testing processors, \boomf{} and \rsd{}, \ourtool{} achieves an average of \cbavetestcondspdp{}$\times$ \spd{} and \cbavetestcondincov{}$\%$ more \tcov{}. 
On average, \ourtool{} achieves a \cbaveallcondspdp{}$\times$ \spd{} and \cbaveallcondcov{}$\%$ more \tcov{} across all five processors. The result demonstrates the comprehensiveness of \ourtool{} on different coverage metrics.}

\section{Integrating the CB Model with Processor Fuzzers}\label{apd:integrate_cb}

Algorithm~\ref{algo_2_full} shows the integration process of \ourtool{}'s trained contextual bandit (CB) model with processor fuzzers. 
\blue{
The inputs are the trained vulnerability list $\mathcal{A}_{\text{v}}$ and coverage lists $\mathcal{A}_{\text{c}}$ with their learned policies $\pi_v$ and $\pi_c$.
it also uses \ccontext{}s $\mathcal{C}$, the check window $\gamma$, and the maximal fuzzing iteration $m$.
The outputs are the achieved \tcov{} $tot\_cov$.}
\ourtool{} starts from the vulnerability list $\mathcal{A}_v$~(Line \red{3}\blue{4}). 
\blue{In \texttt{FuzzVulList}, it samples a test $a_i$ from $\mathcal{A}_v$ according to the policy $\pi_v$ (Line 9).}
\red{It samples a test based on the learned policy $\pi_v$ and monitors its reward $r$.} 
To ensure effectiveness of tests, \ourtool{} evaluates each selected test $a_i$ every $\gamma$ fuzzing iterations, checking whether it yields a \covi{} $r_i(tot\_cov, a_i)$ (Line \red{34}\blue{31}) by measuring the cumulative \icov{} from its mutated variants $f_{cov}(a_i)$. 
If no \covi{}\red{ is observed}, the test is dropped (\red{Line 35}\blue{Lines 32--33}).  
Otherwise, \ourtool{} resets the tracking record $\#$ and continues to monitor the coverage increment\red{ periodically}~(Line \red{37}\blue{35}). 
Once the\red{ vulnerability} list is exhausted \blue{ (Lines 12--13)}, \blue{\texttt{FuzzVulList} returns the current \tcov{} $tot\_cov$ (Line 14).}

\blue{\ourtool{} then swtiches to \clist{}s through \texttt{FuzzCovList} (Line 5).}
\ourtool{}\red{calculates the cumulative \tcov{} $tot\_cov$ and} selects tests \blue{based on the cumulative \tcov{}} from the coverage lists $\mathcal{A}_c = \{\mathcal{A}_c^1,\dots,\mathcal{A}_c^{|\mathcal{C}|}\}$, where $\mathcal{C}$ represents the configured coverage contexts~(Line \red{4}\blue{5}).
\ourtool{} continuously monitors the total coverage $tot\_cov$ achieved by a processor fuzzer and selects tests according to the policy $\pi(\cdot\mid c)$, corresponding to different intervals of coverage contexts (Lines 16-23). 
\blue{For example, if $tot\_cov < c_1$, it samples a test $a_i$ from the coverage list $\mathcal{A}_c^1$ using the policy $\pi_c(\cdot\mid c_1)$ (Lines 17--18); if $c_1 \leq tot\_cov < c_2$, it samples from $\mathcal{A}_c^2$ using the policy $\pi_c(\cdot\mid c_2)$ (Lines 19--20), and so on for the remaining intervals. This allows \ourtool{} to adapt its selection of tests to the current \tcov{}.}
If all curated tests are exhausted, \ourtool{} selects new tests generated by the processor fuzzer to explore the unique design features of the PUT (Line 26).
\blue{After completing at most $m$ iterations, \texttt{FuzzCovList} returns the final total coverage $tot\_cov$ (Line 29).}
\red{For example, if $tot\_cov$ is in the interval $[c_1,c_2)$, \ourtool{} selects tests from the coverage list $\mathcal{A}_c^2$ associated with $c_2$.}

\input{Code/algo2_all}


\red{\section{\red{Distributions of PP-tests from Baseline Fuzzers}}\label{apd:testdist}}
\red{Since \thehuzz{} and \cascade{} apply distinct fuzzing mechanisms, they generate tests that are complementary with each other and help \ourtool{} explore design features that they cannot explore individually. 
We analyze the distributions of the PP-tests from both fuzzers on the coverage lists after training \ourtool{} for $10K$ steps using the minimized corpus and evaluate how the tests enhance coverage on PUTs.}

\red{Figure~\ref{fig:testdist} shows that PP-tests from \cascade{} dominate the coverage lists of $55\%$, $60\%$, and $65\%$ coverage contexts. 
In fact, all tests in the $55\%$ and $60\%$ coverage lists are from \cascade{}, and at least $95\%$ of the tests in the $65\%$ coverage list also originate from \cascade{}.
The results demonstrate that \cascade{} excels at generating tests that efficiently explore broad or commonly executed design spaces.
In contrast, PP-tests from \thehuzz{} contribute a maximum of $6\%$ on the $65\%$ coverage list but account for $100\%$ of the tests in the $70\%$ coverage list.
The results demonstrate that \thehuzz{} is effective at generating tests that explore corner cases of PUTs.}

\red{Overall, the distribution of PP-tests across fuzzers shows the use of fuzzers with distinct mechanisms to construct diverse PP-test corpora for training \ourtool{}'s CB model, thereby improving its ability to explore a wide range of design spaces in PUTs.}

%% file: Code/config_theta.tex
\begin{algorithm}[t]
\caption{Fine-Tune Adaptive Thresholds.}
\label{alg:fine_tune}
\begin{algorithmic}[1]
  \State \textbf{Inputs: $\mathcal{A}_{\text{corpus}}, \mathcal{C}, k, \gamma, n, f$}
  \State \textbf{Outputs: $\theta$}
  \State $\theta \gets \emptyset$, $k_{\mathrm{upper}} = (1+f) * k$, $k_{\mathrm{lower}} = (1-f) * k$  
  \ForAll{$c \in \mathcal{C}$ in reverse order}
      \State $\ell \gets 0.00,\; h \gets 100.00,\; \mathsf{t_c} \gets 0.00$

      \For{$i=1$ to $14$}     \Comment{Binary search}
          \State $m \gets (\ell + h)/2$
          \State $\mathcal{A} \gets \Call{AdapCB}{\mathcal{A}_{\text{corpus}}, c, k, \gamma, m, n}$
          \If{$\mathcal{|A|} > k_{\mathrm{upper}}$} $\ell \gets m$
          \ElsIf{$\mathcal{|A|} < k_{\mathrm{lower}}$} $h \gets m$
          \Else {} $\mathsf{t_c} \gets m; \textbf{break}$ 
          \EndIf
      \EndFor

      \State $\theta \gets \theta \cup \mathsf{\{t_c\}}$
      \State $\mathcal{A} \gets \Call{AdapCB}{c, t_c, \mathcal{A}_{\text{corpus}}, n}$
      \State $\mathcal{A}_{\text{corpus}} \gets \mathcal{A}_{\text{corpus}} \setminus \mathcal{A}$
  \EndFor

  \State \Return $\theta$
\end{algorithmic}
\end{algorithm}

%% file: Table/boom_b2_v3.tex
\def\arraystretch{1.3}
\begin{table}[h]
\caption{Instructions observed to increment the \texttt{minstret} twice.}
\label{tab:tb_boom_b2}
\centering
\resizebox{0.85\columnwidth}{!}{%
\begin{tabular}{cc}
\multicolumn{2}{c}{\xrfill[0.7ex]{1pt} \textbf{  On Both \boomt{} and \boomf{}}  \xrfill[0.7ex]{1pt}}                                                                                                               \\
\textbf{CSR Instructions}                                                             & \begin{tabular}[c]{@{}c@{}}\texttt{CSRRC}, \texttt{CSRRCI}, \texttt{CSRRS}, \texttt{CSRRSI},\\ \texttt{CSRRW}, \texttt{CSRRWI}\end{tabular}   \\ \hline
\textbf{\begin{tabular}[c]{@{}c@{}}Memory\\ Synchronization\end{tabular}}             & \texttt{FENCE}, \texttt{FENCE.I}                                                                          \\ \hline
\textbf{Load \& Store}                                                                 & \begin{tabular}[c]{@{}c@{}}\texttt{LB}, \texttt{LBU}, \texttt{LH}, \texttt{LHU}, \texttt{LW}, \texttt{LWU}, \texttt{LD},\\ \texttt{SB}, \texttt{SH}, \texttt{SW}, \texttt{SD}\end{tabular} \\ \hline
\textbf{\begin{tabular}[c]{@{}c@{}}Load-Reserved \& \\ Store-Conditional\end{tabular}} & \texttt{LR.W}, \texttt{LR.D}, \texttt{SC.W}, \texttt{SC.D}                                                                  \\ 
\multicolumn{2}{c}{\xrfill[0.7ex]{1pt} \textbf{  On \boomf{} Only} \xrfill[0.7ex]{1pt}}                                                                                                                                   \\
\textbf{Branch}                                                                       & \texttt{JALR}, \texttt{BGE}, \texttt{BLT}, \texttt{BLTU}, \texttt{BNE}                                                               \\ \hline
\textbf{Multiplication}                                                               & \begin{tabular}[c]{@{}c@{}}\texttt{MUL}, \texttt{MULH}, \texttt{MULHSU}, \texttt{MULHU},\\ \texttt{MULW}\end{tabular}                \\ \hline
\end{tabular}%
}
\end{table}

%% file: Code/algo2_all.tex
\begin{algorithm}[H]
\begin{minipage}{1\linewidth}
\caption{Integrating the CB Model with Processor Fuzzers}\label{algo_2_full}
\begin{algorithmic}[1]
    \State \blue{\textbf{Inputs: $\mathcal{A}_{v},\pi_v,\mathcal{A}_{c},\pi_c,\mathcal{C}, \gamma, m$}}
    \State \blue{\textbf{Outputs: $tot\_cov$}}
\Function{Main}{}
    \State $tot\_cov \gets$ \Call{FuzzVulList}{$\mathcal{A}_v,\pi_v,\gamma$}
    \State $tot\_cov \gets$ \Call{FuzzCovList}{$\mathcal{A}_c,\pi_c,\mathcal{C},\gamma,tot\_cov,m$}
    \State \Return $tot\_cov$
\EndFunction

\Function{FuzzVulList}{$\mathcal{A}_v$, $\pi_v$, $\gamma$}
    \While{true}
        \State $a_i\sim \pi_v(\cdot)$
        \State \textsc{DropTest}$(a_i, \#(a_i), r(tot\_cov, a_i),f_{\mathrm{cov}}(a_i),$ $\mathcal{A}_v, \pi_v, \gamma)$
        \State $tot\_cov \gets tot\_cov + f_{cov}(a_i)$
        \If{$|\mathcal{A}_v| = 0$}
            \State \textbf{break}
        \EndIf
    \EndWhile
    \State \Return $tot\_cov$
\EndFunction

\Function{FuzzCovList}{$\mathcal{A}_c$, $\pi_c$, $\mathcal{C}$, $\gamma$, $tot\_cov$, $m$}
    \For{$i = 1, 2, \dots, m$}
        \If{$tot\_cov < c_1$}
            \State $a_i\sim\pi_c(\cdot\mid c_1)$, $a_i\in\mathcal{A}_c^1$
        \ElsIf{$c_1 \leq tot\_cov < c_2$}
            \State $a_i\sim\pi_c(\cdot\mid c_2)$, $a_i\in\mathcal{A}_c^2$
        \Statex\hspace{\algorithmicindent}...
        \setcounter{ALG@line}{24}
        \ElsIf{no arm $a_i$ for $tot\_cov$}
            \State Let fuzzer generate seed $a_i$
        \EndIf
        \State $tot\_cov \gets tot\_cov + f_{cov}(a_i)$
        \State \Call{DropTest}{$a_i, \#(a_i), r(tot\_cov, a_i), f_{\mathrm{cov}}(a_i),$ $\mathcal{A}_c, \pi_c, \gamma$}
    \EndFor
    \State \Return $tot\_cov$
\EndFunction

\Function{DropTest}{$a$, $\#$, $r$, $f_{\mathrm{cov}}$, $\mathcal{A}$, $\pi$, $\gamma$}
    \State $\# \gets \#+ 1 $; $r \gets r + f_{\mathrm{cov}}$
    \If{$\# \geq \gamma$ and $r = 0$}
        \State $\mathcal{A} \gets \mathcal{A}\setminus \{a\}$
    \Else
        \State $r \gets 0$; $\# \gets 0$
    \EndIf
    \State \Return $\#$, $r$, $\mathcal{A}$, $\pi$
\EndFunction

\end{algorithmic}
\end{minipage}
\end{algorithm}